\documentclass{aa}

\usepackage{graphicx}
\usepackage{txfonts}
\usepackage[colorlinks=true,citecolor=blue,linkcolor=blue]{hyperref}
\usepackage{cleveref}
\usepackage{siunitx}
\usepackage[draft]{changes}
\usepackage[all]{nowidow}

\DeclareSIUnit \parsec 	    {pc}
\DeclareSIUnit \eV 			{eV}
\DeclareSIUnit \keV 		{keV}
\DeclareSIUnit \Msun 		{M_\odot}

\definechangesauthor[name={Krut}, color=olive]{ak}
\definechangesauthor[name={Arguelles}, color=orange]{ca}
\definechangesauthor[name={Rueda}, color=purple]{jar}

\begin{document} 

   \title{The geodesic motion of S2 and G2 as a test of the fermionic dark matter {nature} of our galactic core}

   \author{E.~A.~Becerra-Vergara
          \inst{1}\fnmsep\inst{2}\fnmsep\inst{3}
          \and
          C.R.~Arg\"uelles\inst{1}\fnmsep\inst{2}\fnmsep\inst{4}
          \and
          A.~Krut\inst{1}\fnmsep\inst{2}
          \and
          J.~A.~Rueda\inst{1}\fnmsep\inst{2}\fnmsep\inst{5}\fnmsep\inst{6}\fnmsep\inst{7}
          \and
          R.~Ruffini\inst{1}\fnmsep\inst{2}\fnmsep\inst{5}\fnmsep\inst{6}\fnmsep\inst{8}
          }
   \institute{ICRANet, Piazza della Repubblica 10, I--65122 Pescara, Italy\\
   \email{eduar.becerra@icranet.org,andreas.krut@icranet.org, jorge.rueda@icra.it, ruffini@icra.it}
           \and
        ICRA, Dipartimento di Fisica, Sapienza Universit\`a di Roma, P.le Aldo Moro 5, I--00185 Rome, Italy
         \and
    Grupo de Investigaci\'on en Relatividad y Gravitaci\'on, Escuela de F\'isica, Universidad Industrial de Santander, A. A. 678, Bucaramanga 680002, Colombia
        \and
        Facultad de Ciencias Astron\'omicas y Geof\'isicas, Universidad Nacional de La Plata, Paseo del Bosque, B1900FWA La Plata, Argentina\\
        \email{carguelles@fcaglp.unlp.edu.ar}
        \and
        ICRANet-Ferrara, Dipartimento di Fisica e Scienze della Terra, Universit\`a degli Studi di Ferrara, Via Saragat 1, I--44122 Ferrara, Italy
        \and
        Dipartimento di Fisica e Scienze della Terra, Universit\`a degli Studi di Ferrara, Via Saragat 1, I--44122 Ferrara, Italy
        \and
        INAF, Istituto de Astrofisica e Planetologia Spaziali, Via Fosso del Cavaliere 100, 00133 Rome, Italy
        \and
        INAF, Viale del Parco Mellini 84, 00136 Rome  Italy
        }

   \date{Received Month day, year; accepted Month day, year}

 
\abstract{
The S-stars motion around the Galactic center implies that the central gravitational potential is dominated by a compact source, Sagittarius A* (Sgr~A*), with a mass of {about} $\SI{4E6}{\Msun}$, traditionally assumed to be a massive black hole (BH). {Particularly important for this hypothesis, and for any alternative model, is the explanation of the multiyear, accurate astrometric data of the S2 star around Sgr~A*, including the relativistic redshift which has been recently verified. Another relevant object is G2, whose most recent observational data challenge the massive BH scenario}: its post-pericenter radial velocity is lower than the expectation from a Keplerian orbit around the putative massive BH. This scenario has traditionally been reconciled by introducing a drag force on G2 by an accretion flow. {Alternatively to the central BH scenario, we here demonstrate that the observed motion of both S2 and G2 is explained in terms of the \textit{\mbox{dense~core~--~diluted~halo}} fermionic dark matter (DM) profile, obtained from the fully relativistic Ruffini-Arg\"uelles-Rueda (RAR) model. It has been already shown that for fermion masses \SIrange{48}{345}{\kilo\eV}, the RAR-DM profile accurately fits the rotation curves of the Milky Way halo. We here show that the solely gravitational potential of such a DM profile, for a fermion mass of \SI{56}{\kilo\eV}, explains: 1) all the available time-dependent data of the position (orbit) and {line-of-sight} radial velocity (redshift function $z$) of S2; 2) the combination of the special and general relativistic redshift measured for S2; 3) the currently available data on the orbit and $z$ of G2; and 4) its post-pericenter passage deceleration without introducing a drag force. For both objects, we find that the RAR model fits better the data than the BH scenario: the mean of reduced chi-squares of the time-dependent orbit and $z$ data are, for S2, $\langle\bar{\chi}^2\rangle_{\rm S2, RAR}\approx 3.1$ and $\langle\bar{\chi}^2\rangle_{\rm S2, BH}\approx 3.3$ and, for G2, $\langle\bar{\chi}^2\rangle_{\rm G2, RAR}\approx 20$ and $\langle\bar{\chi}^2\rangle_{\rm G2, BH}\approx 41$. If we look at the fit of the corresponding $z$ data, while for S2 we find comparable fits, i.e, $\bar{\chi}^2_{z,\rm RAR}\approx 1.28$ and $\bar{\chi}^2_{z,\rm BH}\approx 1.04$, for G2 only the RAR model can produce an excellent fit of the data, i.e. $\bar{\chi}^2_{z,\rm RAR}\approx 1.0$ and $\bar{\chi}^2_{z,\rm BH}\approx 26$.
In addition,} the critical mass for gravitational collapse of a {degenerate \SI{56}{\kilo\eV}-fermion DM} core into a BH is $\sim \SI{E8}{\Msun}$. {This result may provide the initial seed for the} formation of the observed central supermassive BH in active galaxies, such as M87.
}
   
\titlerunning{S2 and G2 as test of fermionic dark matter}
\authorrunning{Becerra-Vergara, et al.}

\maketitle
%

\section{Introduction}\label{sec:1}

The monitoring over the last decades of the motion of the so-called S-stars near the Galactic center has revealed that the gravitational potential in which they move is dominated by a massive compact source at the center, Sagittarius A* (Sgr~A*) \citep{2009ApJ...707L.114G, 2017ApJ...837...30G}. The S-star dynamics implies a mass for Sgr~A* of $\approx \SI{4.1E6}{\Msun}$, traditionally associated in the literature with a massive black hole (BH) \citep{2018A&A...618L..10G, 2008ApJ...689.1044G, 2010RvMP...82.3121G}.

Most interesting among the objects moving near and around Sgr~A* are S2 and G2. The star S2 describes an elliptical orbit with focus on Sgr~A*, a period of $16.05$~yr and the second closest pericenter among the S-stars, $r_{p(S2)}\approx \SI{0.6}{m\parsec}$ \citep{2009ApJ...707L.114G,2017ApJ...837...30G}. The S2 orbit constrains best the Sgr~A* mass, but its pericenter at $\sim 1500\,r_{\rm Sch}$ from Sgr~A*, is too far for univocally infer a putative massive BH of Schwarzschild radius $r_{\rm Sch} = 2 G M_{\rm BH}/c^2$, being $M_{\rm BH}$ its mass.

The most recent measurements of the motion of G2 after the peripassage around Sgr~A* represent a further challenge for the massive BH hypothesis. The G2 radial velocity is lower than the one from a Keplerian motion around the massive BH, which has been reconciled by introducing the action of a drag force exerted by an accretion flow \citep{2017ApJ...840...50P,2019ApJ...871..126G}.

Our aim here is to show that, instead, the \textit{dense~core~--~diluted~halo} {DM} density distribution of a general relativistic system of \SI{56}{\kilo\eV} fermions, following the extended Ruffini-Argüelles-Rueda (RAR) model \citep{2018PDU....21...82A,2019PDU....24100278} explains, without invoking the massive BH or a drag force, both the S2 and G2 orbits. We shall make use of the most complete data of the S2 orbit over the last $26$~yr {\citep{2017ApJ...837...30G,2018A&A...615L..15G}, including the recent data released by  \citet{Do664}}, and the $4$-year data of the G2 motion after its pericenter passage \citep{2019ApJ...871..126G}.
\section{The RAR model of dark matter}\label{sec:2}

\begin{figure}
    \centering
	\includegraphics[width=\hsize,clip]{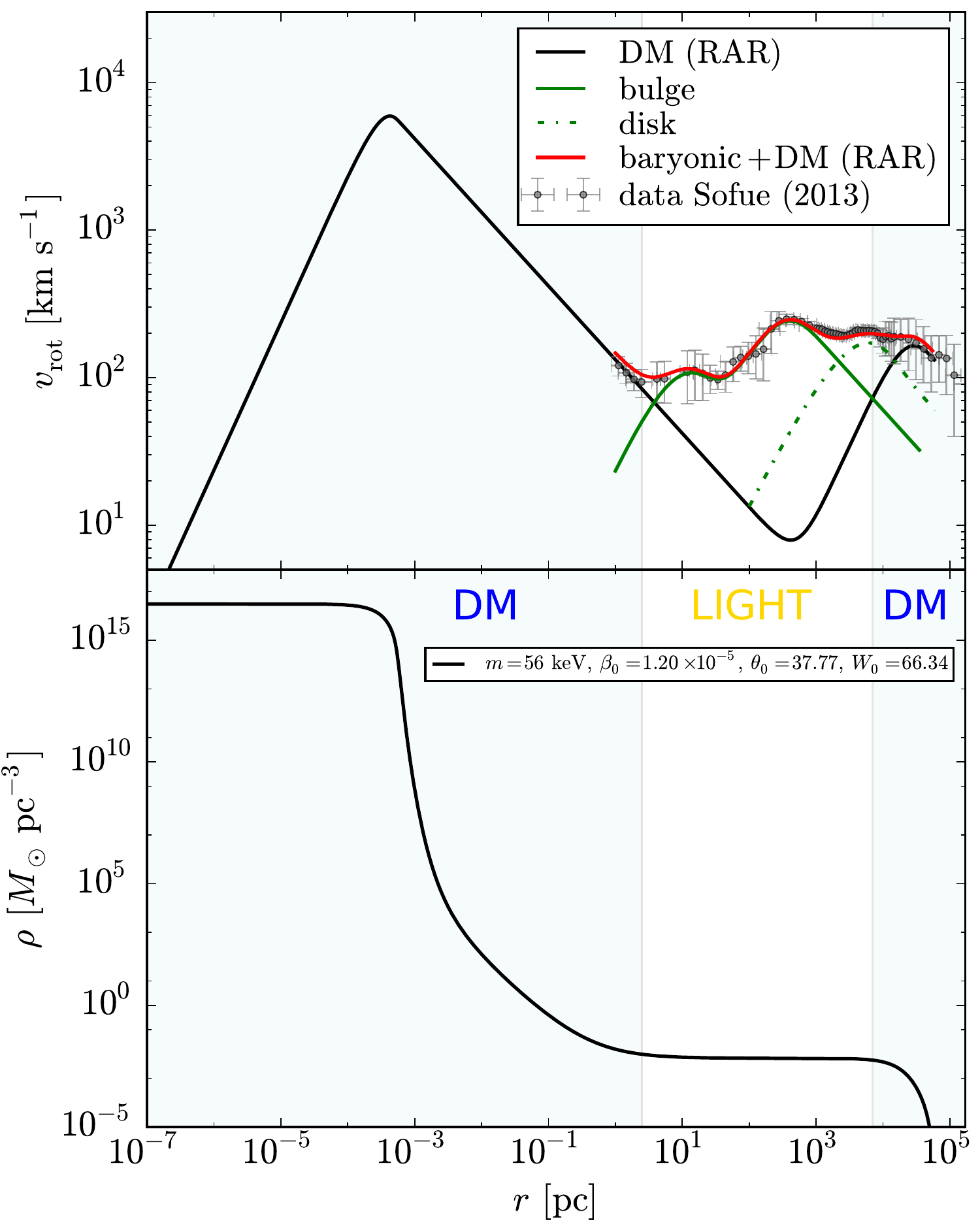}
	\caption{Milky Way rotation curve and DM density {profile from the extended-RAR model with a core mass of $M_c = M(r_c)=\SI{3.5E6}{\Msun}$}. Top: DM (black) and baryonic (bulge + disk) contribution to the rotation curve $v_{\rm rot}$ (total in red). Bottom: DM density profile. The baryonic model and the data are taken from \citep{2013PASJ...65..118S}. The parameters of the extended-RAR model in this case are: fermion mass $m c^2=\SI{56}{\kilo\eV}$, temperature parameter {$\beta_0=1.1977\times 10^{-5}$, degeneracy parameter $\theta_0=37.7656$ and energy cutoff parameter $W_0=66.3407$}. For the RAR model fitting of the Milky Way we follow \citep{2018PDU....21...82A}; see also \Cref{sec:SM1}.}\label{fig:vrot}
\end{figure}

The RAR model equilibrium equations consist of the Einstein equations in spherical symmetry for a perfect fluid energy-momentum tensor, with pressure and density given by Fermi-Dirac statistics and closure relations determined by the Klein and Tolman conditions of thermodynamic equilibrium \citep{2015MNRAS.451..622R}. The solution to this system of equations leads to a continuous and novel \textit{dense~core~--~diluted~halo} DM profile from the center all the way to the galactic halo\footnote{Similar \textit{core-halo} profiles with applications to fermionic DM were also obtained in \citep{2002PrPNP..48..291B}, and more recently in \citep{2015PhRvD..92l3527C} from a statistical approach within Newtonian gravity.} { \citep[see][for its applications]{2015ARep...59..656S,2016JCAP...04..038A,2017IJMPD..2630007M}.} This corresponds to the original version of the RAR model, with a unique family of density profile solutions that behave as $\rho(r) \propto r^{-2}$, at large radial distances from the center. This treatment was extended in \citep{2018PDU....21...82A} (see \Cref{sec:SM1}) by introducing in the distribution function (DF) a cutoff in momentum space (i.e. accounting for particle-escape effects) that allows to define the galaxy border. Such RAR model extension was successfully applied to explain the Milky Way rotation curve as shown in \Cref{fig:vrot}, implying a more general \textit{dense~core~--~diluted~halo} behavior for the DM distribution as follows:
\begin{itemize}
\item
A DM core with radius $r_c$ (defined at the first maximum of the twice-peaked rotation curve), whose value is shown to be inversely proportional to the particle mass $m$, in which the density is nearly uniform. This central core is supported against gravity by the fermion degeneracy pressure and general relativistic effects are appreciable.
\item
Then, there is an intermediate region characterized by a sharply decreasing density where quantum corrections are still important, followed by an extended and diluted plateau. This region extends until the halo scale-length $r_h$ is achieved (defined at the second maximum of the rotation curve).
\item
Finally, the DM density reaches a Boltzmannian regime supported by thermal pressure with negligible general relativistic effects, and showing a behavior $\rho\propto r^{-n}$ with $n>2$ due to the phase-space distribution cutoff which leads to a DM halo bounded in radius (i.e. $\rho\approx 0$ occurs when the particle escape energy approaches zero).
\end{itemize}

As it has been explicitly shown in \citet{2019IJMPD..2843003A,2019PDU....24100278,2018PDU....21...82A}, this kind of \textit{dense~core~--~diluted~halo} density profile suggests that the DM could explain both the mass of the dark compact object {in Sgr A*} as well as the one of the halo. It applies not only to the Milky Way but also in other galactic structures from dwarfs to ellipticals to galaxy clusters \citep{2019PDU....24100278}. Specifically, the Milky Way analysis \citep{2018PDU....21...82A} has shown that indeed this DM profile can explain the dynamics of the closest S-cluster stars (including S2) around Sgr~A*, all the way to the halo rotation curve without spoiling the baryonic bulge-disk components. The analysis of the S-stars was there made through a simplified circular velocity analysis in general relativity, {constraining the allowed fermion mass to $mc^2\approx \SIrange{50}{345}{\kilo\eV}$}. Here, we extend such an analysis by making a full reconstruction of the object's geodesic in full general relativity, and apply it to S2 and G2. \Cref{fig:vrot} shows the DM density profile and its contribution to the rotation curve for the Milky Way for \SI{56}{\kilo\eV} DM fermions.
%

\section{Orbit and radial velocity of S2 and G2}\label{sec:3}

{For obtaining the S2 or G2 positions (orbit) and the corresponding line-of-sight radial velocity (i.e. the redshift function; see \Cref{sec:SM2}) at each time, we solve the equations of motion for a test particle (see \Cref{sec:SM3}) in the gravitational field produced by:}
\begin{enumerate}
\item
A central Schwarzschild massive BH. { \citet{2018A&A...615L..15G} reported a BH mass of $M_{\rm BH} = \SI{4.1E6}{\Msun}$ from the fit of the most recent measurements of the position and velocity of S2. A more recent analysis by \citet{Do664}, reported a BH mass of \SI{3.975E6}{\Msun}. Those works use a second-order post-Newtonian ({2PN}) model to describe the object's motion. In order to compare and contrast the BH and the DM-RAR hypotheses on the same ground, namely using the same analysis method and treatment, we perform our own fit of the data for the BH case using a full general relativistic modeling by solving the equations of motion in the Schwarzschild metric (see \Cref{sec:SM3}). From our analysis of S2, we obtain very close (but not equal) model parameters to the ones presented in \citet{2018A&A...615L..15G} and \citet{Do664}; see \Cref{tab:parameter}. In particular, we obtain a BH mass of $M_{\rm BH} = \SI{4.075E6}{\Msun}$}.
\item 
A {fermionic} DM distribution obtained from the extended-RAR model; see \Cref{sec:SM1}. {As it is shown in \citet{2018PDU....21...82A}, the fermion mass must be larger than \SI{48}{\kilo\eV} and lower than \SI{345}{\kilo\eV}}. We here present the results {of the solution of the equations of motion in the metric produced by the DM distribution of \SI{56}{\kilo\eV}-fermions, with corresponding RAR model parameters as shown in \Cref{fig:vrot}. We obtain an excellent fit of the data for a mass of the DM \textit{quantum-core}, $M_c \equiv M(r_c) = \SI{3.5E6}{\Msun}$; see \Cref{tab:parameter}}.
\end{enumerate}

{
It has been previously reported that the BH mass, $M_{\rm BH}$, and the Galactic center distance, $D_{\odot}$, show some correlation \citep{2018A&A...615L..15G,Do664}. We here adopt the distance to the Galactic center as a fixed parameter, $D_{\odot} = \SI{8}{k\parsec}$. Instead, as we have mentioned, for $M_{\rm BH}$ we seek for a best-fit value. Thus, in principle, not considering together $D_\odot$ and $M_{\rm BH}$ as adjustable parameters might have some impact on the inferred values. However, as it can be seen from \Cref{tab:parameter}, our inferred values for the parameters of the BH model agree with the ones reported in previous analyses, including the BH mass, see e.g. \citet{2018A&A...615L..15G,Do664}.
}

{Due to the regular initial condition applied to solve the extended RAR model equations, i.e. $\rho(r=0)=$const. (see \Cref{fig:vrot} and \Cref{sec:SM1} for details), the DM \textit{quantum-core} is not directly comparable with a BH, which is characterized by a central singularity. However, it is possible to compare the responsible mass of the innermost Keplerian behavior (i.e. power law $\propto r^{-1/2}$ in the velocity curve) of orbiting objects in both scenarios. In the RAR-model case, the Keplerian behavior arises just outside the core radius (see  \Cref{fig:vrot}). The corresponding `Keplerian mass', say $M_K$, describing the Keplerian trend is slightly larger than the DM core mass $M_c$, due the slight mass contribution along the sharp density drop. For larger radii already in the diluted plateau density, the mass contribution to $M_K$ is negligible, up to the ending Keplerian trend occurring at about few $\SI{1E2}{\parsec}$ (curiously at the peak of the bulge velocity curve, see  \Cref{fig:vrot}). For a \textit{quantum-core} mass of $M_c = \SI{3.5E6}{\Msun}$, we find the corresponding Keplerian mass $M_K = \SI{4.048E6}{\Msun}$. This value is indeed very close the one inferred for the BH scenario, $M_{\rm BH} = \SI{4.075E6}{\Msun}$, and should be kept in mind (besides $M_c$) when comparing both models regarding the (stellar) dynamics in the surroundings of Sgr~A*.}

\begin{table*}[htbp!]
\caption{{Summary of the {inferred} best-fit values of the model {and the (osculating) orbital} parameters for S2 and G2 within the RAR model {(fermion mass $\SI{56}{\kilo\eV}$, DM core mass $M_c = \SI{3.5E6}{\Msun}$)} and the massive BH model {(BH mass $M_{\rm BH}=\SI{4.075E6}{\Msun}$)}. We refer to \Cref{sec:SM3} for details on the {definition of the parameters and on the} fitting procedure.}}\label{tab:parameter}
\centering
\scriptsize{
\begin{tabular}{lcccccccccc}
\hline
\multicolumn{1}{c}{\bf{Parameter}} &  &  & \multicolumn{3}{c}{\bf{S2}}         &  &  & \multicolumn{3}{c}{\bf{G2}}           \\ \cline{4-6} \cline{9-11}
                           &  &  & \bf{RAR}      &  & \bf{BH }              &  &  & \bf{RAR}      &  & \bf{BH}                \\ \cline{1-1} \cline{4-4} \cline{6-6} \cline{9-9} \cline{11-11}  
Semimajor Axis, $a$~({as}) &  &  & {$0.1252$}&  & {$0.1252$}  &  &  & {$1.0960$}  &  & {$1.1941$}  \\
Eccentricity, $e$         &  &  & {$0.8866$} &  & {$0.8863$}  &  &  & {$0.9823$}  &  & {$0.9853$}  \\
Distance to Pericenter, $r_p$~({as}) &  &  & {$0.0142$} &  & {$0.0143$}  &  &  & {$0.0194$} &  & {$0.0180$} \\
Distance to Apocenter, $r_a$~({as}) &  &  & {$0.2361$} &  & {$0.2362$} &  &  & {$2.1725$} &  & {$2.3701$}  \\
Argument of Pericenter, $\omega$~($^{\circ}$) &  &  & {$66.7724$} & & {$66.4697$} &  &  & {$81.8391$}  &  & {$82.0001$}  \\
Inclination, $i$~($^{\circ}$)&  &  & {$134.3533$} &  & {$134.3505$} &  &  & {$121.8993$}  &  & {$119.1000$}  \\
Ascending Node, $\Omega$~($^{\circ}$) &  &  & {$228.0240$} &  & {$227.9681$} &  &  & {$50.8398$}  &  & {$50.7782$}  \\
{$X_0$ (mas)} &  &  & {$-0.1557$} &  & {$-0.0830$} &  &  & {$0.0248$} &  & {$0.0251$}\\
{$Y_0$ (mas)} &  &  & {$2.5527$} &  & {$2.4893$} &  &  & {$-0.0160$} &  & {$-0.0140$}\\
Orbital Period, $P$~($yr$) &  &  & {$16.0539$} &  & {$16.0506$} &  &  & {$416.3400$} &  & {$470.1610$}\\
\hline
{$\bar{\chi}^2_X$} &  &  & {$1.5964$}  &  & {$1.8004$} &  &  & {$33.3339$} &  & {$83.9950$}\\
{$\bar{\chi}^2_Y$} &  &  & {$6.3411$}  &  & {$7.2332$} &  &  & {$26.8419$} &  & {$11.2646$}\\
{$\bar{\chi}^2_z$} &  &  & {$1.2799$}  &  & {$1.0421$} &  &  & {$0.9960$} &  & {$26.3927$}\\
{$\langle \bar{\chi}^2\rangle$} &  &  & {$3.0725$}  &  & {$3.3586$} &  &  & {$20.3906$} &  & {$40.5507$}\\
\hline
\end{tabular}
}
\end{table*}

We present in \Cref{sec:SM3} the equations of motion for the general spherically symmetric metric and the procedure we use to fit the observational data of the apparent orbit and line-of-sight radial velocity (i.e. the redshift function) in both scenarios.

\begin{figure*}[ht!]%
	\centering%
	\includegraphics[width=\hsize,clip]{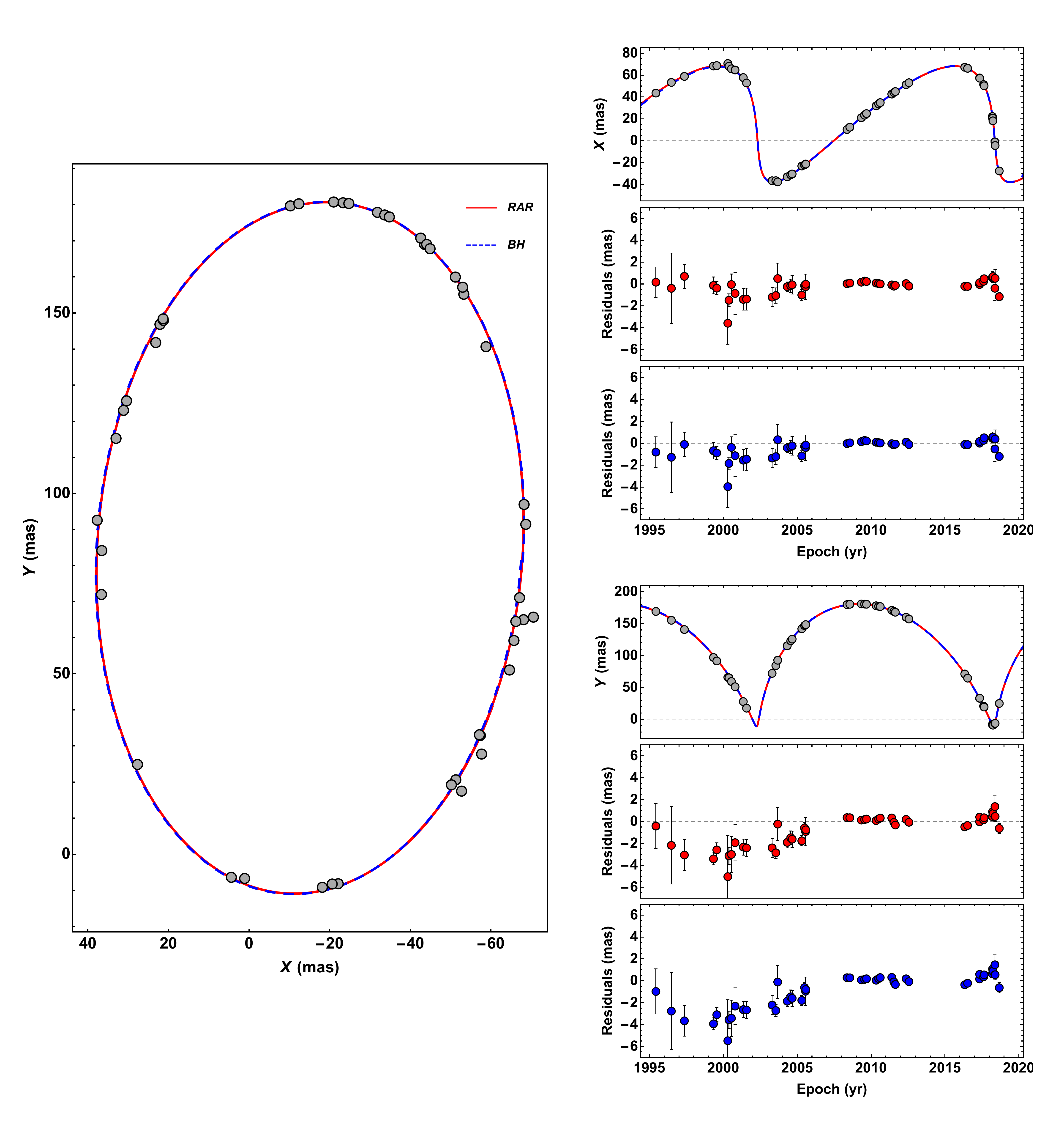}
	\caption{Theoretical and observed orbit of S2 around Sgr~A*. {The left panel shows the orbit ($X$ vs $Y$) and the right panel shows the $X$ and $Y$ position as a function of time, and the respective residuals of the best-fit for each model. The theoretical models are} calculated by solving the equations of motion of a test particle in the gravitational field of: 1) a Schwarzschild BH of \SI{4.075E6}{\Msun} (blue dashed curves), and 2) the DM distribution obtained from the extended RAR model for \SI{56}{\kilo\eV}-fermions (red curves). The mass of the quantum core in the RAR model {is \SI{3.5E6}{\Msun}}. \Cref{tab:parameter} shows the parameters of each model. We use the observational data reported in \citet{Do664}.}\label{fig:S2}%
\end{figure*}

\begin{figure*}%
	\centering%
	\includegraphics[width=0.8\hsize,clip]{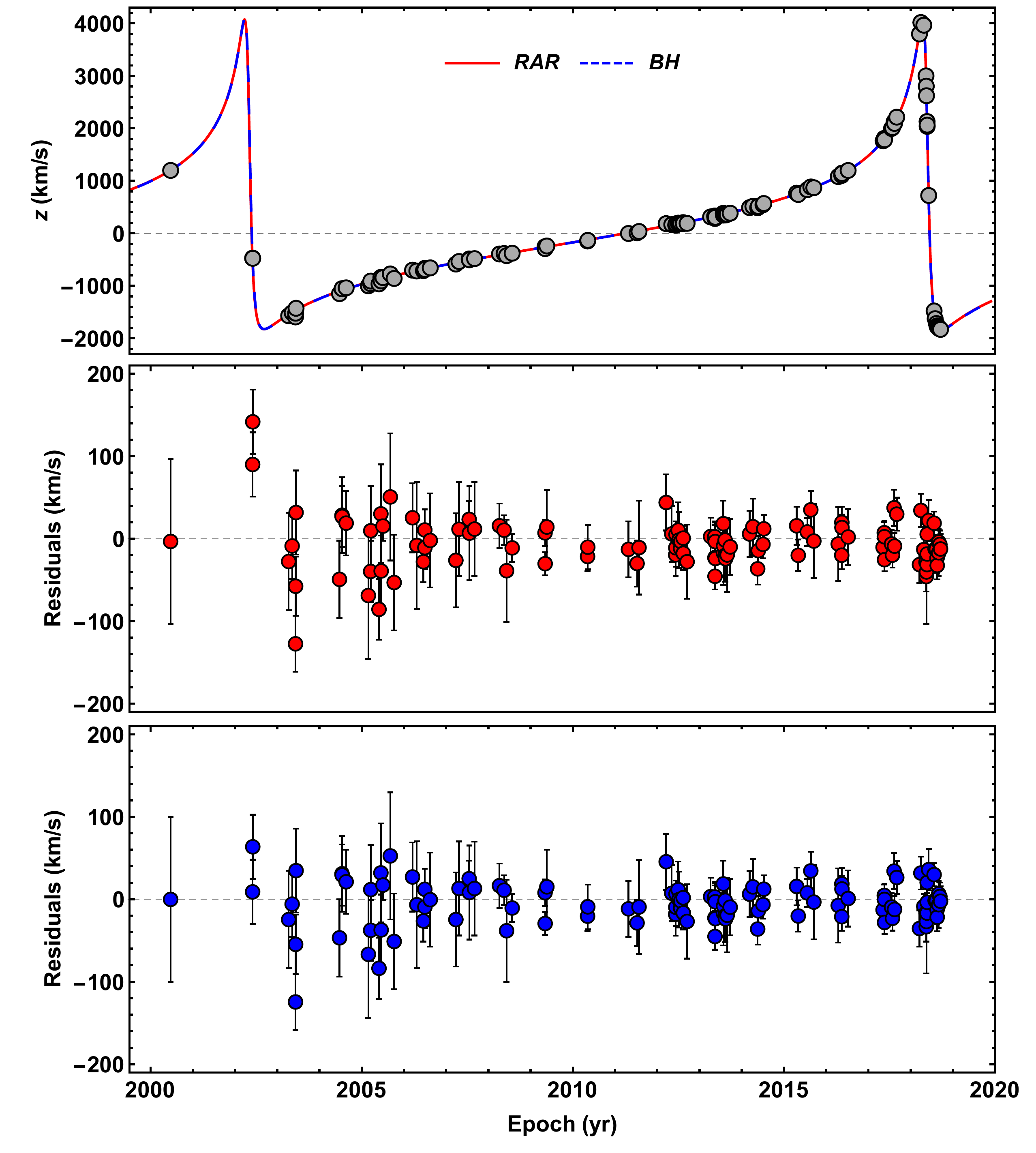}
	\caption{Theoretical and observed {line-of-sight radial velocity (i.e. the redshift function $z$; see \Cref{sec:SM2}) of S2. The theoretical models are} calculated by solving the equations of motion of a test particle in the gravitational field of: 1) a Schwarzschild BH of \SI{4.075E6}{\Msun} (blue dashed curves), and 2) the DM distribution obtained from the extended RAR model for \SI{56}{\kilo\eV}-fermions (red curves). The mass of the quantum core in the RAR model {is \SI{3.5E6}{\Msun}}. \Cref{tab:parameter} shows the parameters of each model. We use the observational data reported in \citet{Do664}.}\label{fig:S2z}%
\end{figure*}

{\Cref{fig:S2} shows the results of the above two theoretical scenarios and how they compare with the observational data of the orbit {(observed right ascension, $X$, and declination, $Y$)} for the case of S2. The comparison with the data of the line-of-sight radial velocity is shown in \Cref{fig:S2z}. It is already noticeable by visual inspection of the residuals that both theoretical models can explain the observational data for the orbit with similar accuracy. In fact, the reduced-$\chi^2$ of the model data fit of the S2 radial velocity ($\bar{\chi}^2_z$) and orbit ($\bar{\chi}^2_X$ and $\bar{\chi}^2_Y$), lead to a comparable mean for both scenarios (with some preference for the RAR model): $\langle \bar{\chi}^2 \rangle_{\rm RAR} \approx 3.072$, $\langle \bar{\chi}^2 \rangle_{\rm BH} \approx 3.359$. We refer to \Cref{tab:parameter} for the model parameters and to \Cref{sec:SM3} for details on the fitting procedure.}

\begin{figure*}
	\centering%
	\includegraphics[width=\hsize,clip]{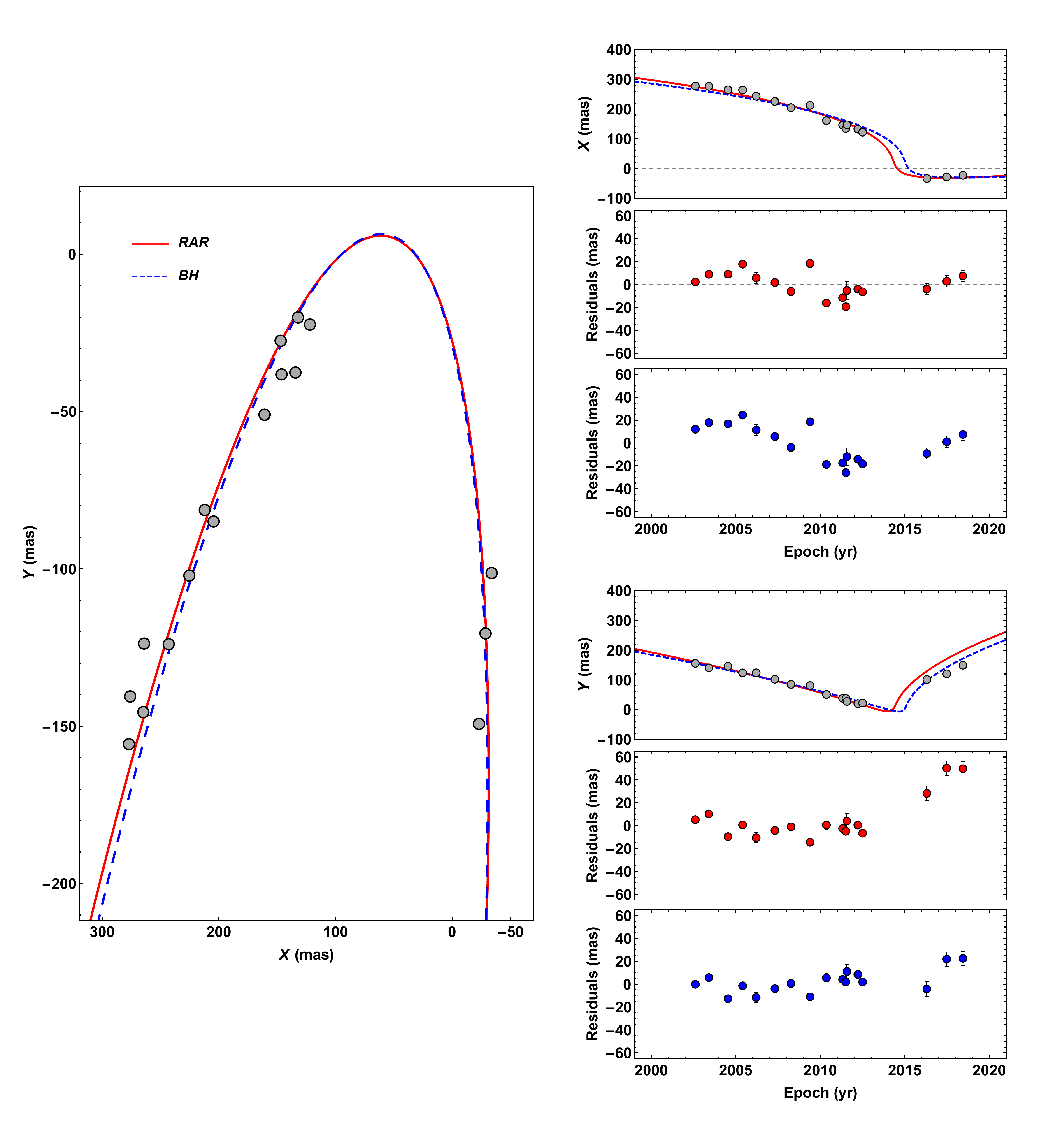}
	\caption{Theoretical and observed orbit of G2 around Sgr~A*. {The left panel shows the orbit ($X$ vs $Y$) and the right panel shows the $X$ and $Y$ position as a function of time, and the respective residuals of the best-fit for each model. The theoretical models are} calculated by solving the equations of motion of a test particle in the gravitational field of: 1) a Schwarzschild BH of \SI{4.075E6}{\Msun} (blue dashed curves), and 2) the DM distribution obtained from the extended RAR model for \SI{56}{\kilo\eV}-fermions (red curves). The mass of the quantum core in the RAR model {is \SI{3.5E6}{\Msun}}. \Cref{tab:parameter} shows the parameters of each model. The observational data has been taken from {\citet{phifer2013keck,2017ApJ...840...50P,2019ApJ...871..126G}}.}\label{fig:G2}%
\end{figure*}

\begin{figure*}
	\centering%
	\includegraphics[width=0.8\hsize,clip]{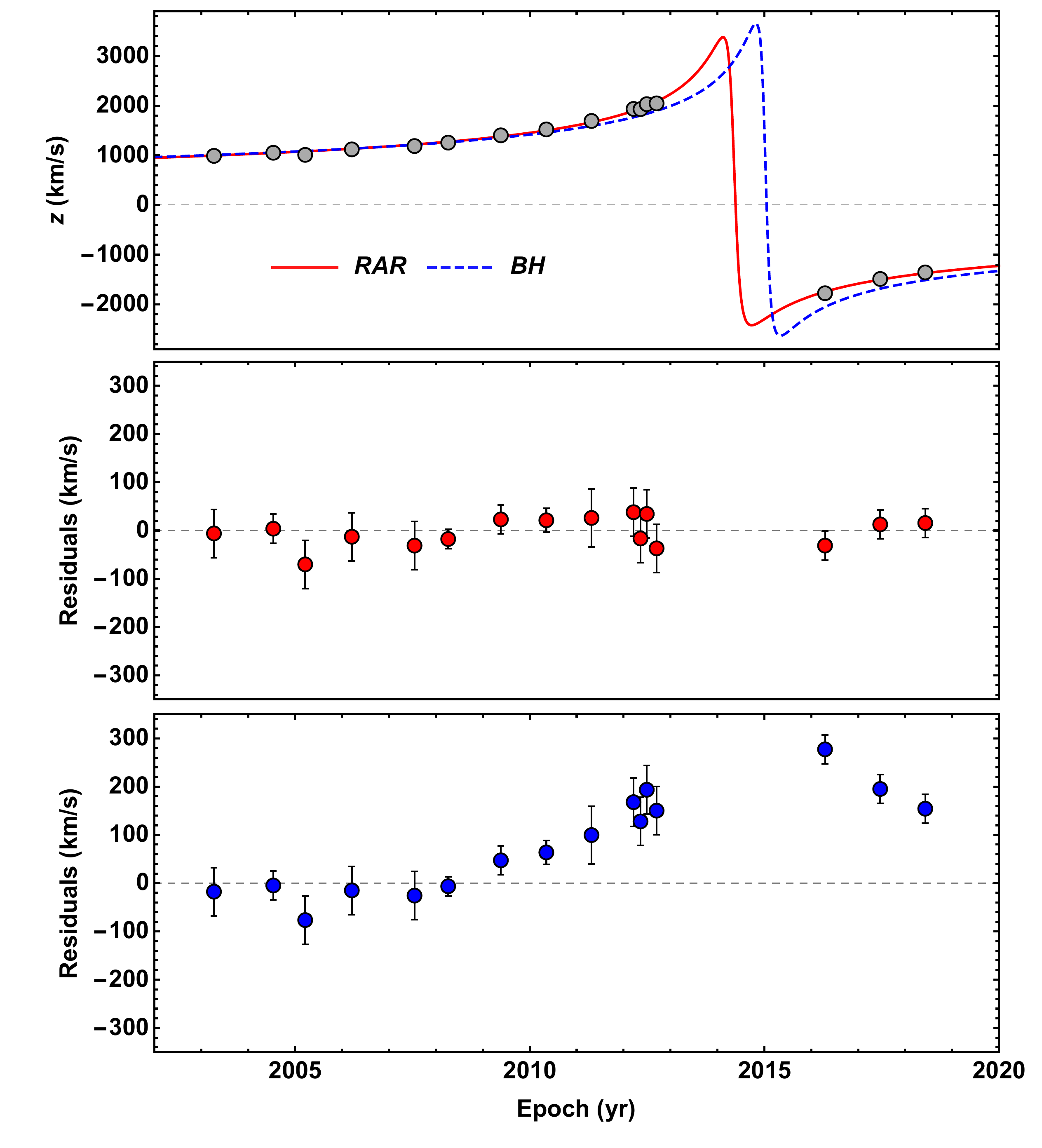}
	\caption{Theoretical and observed {line-of-sight radial velocity (i.e. the redshift function $z$; see \Cref{sec:SM2}) of G2. The theoretical models are} calculated by solving the equations of motion of a test particle in the gravitational field of: 1) a Schwarzschild BH of \SI{4.075E6}{\Msun} (blue dashed curves), and 2) the DM distribution obtained from the extended RAR model for \SI{56}{\kilo\eV}-fermions (red curves). The mass of the quantum core in the RAR model {is \SI{3.5E6}{\Msun}}. \Cref{tab:parameter} shows the parameters of each model. The observational data has been taken from {\citet{phifer2013keck,2017ApJ...840...50P,2019ApJ...871..126G}}}\label{fig:G2vel}%
\end{figure*}

The situation becomes even more interesting in the analogous analysis made for G2. As already shown in \citet{2017ApJ...840...50P,2019ApJ...871..126G}, the G2 orbit shows a radial velocity slower than the one predicted by the geodesic motion in the gravitational field of the massive BH. {Thus,} it has been {there} proposed that G2 is being {slowed} down by a drag force {caused by} an accretion flow onto the massive BH over which G2 should move. The novel major result is that a geodesic in the gravitational field of the DM profile of the extended RAR model naturally predicts such a slowing down (see \Cref{fig:G2,fig:G2vel} and \Cref{tab:parameter}). The higher G2 deceleration is because it moves in the gravitational field produced by the spatially-varying mass profile of the fermionic DM. The above effect of deceleration is instead negligible in the case of S2 due to the shape of the orbit, more precisely due to its size. From its pericenter at {$\sim \SI{0.6}{m\parsec}$} to apocenter at {$\sim \SI{10}{m\parsec}$} (see \Cref{tab:parameter}), S2 moves only a short distance in which the density of the fermionic DM varies considerably less than in the G2 case. Indeed, the orbit of G2, from its pericenter at {$\sim \SI{0.8}{m\parsec}$} to its apocenter at {$\sim \SI{85}{m\parsec}$}, crosses a much larger region where the {DM} density drastically drops off from $\sim \SI{1E15}{\Msun/\parsec}^{-3}$ to $\sim \SI{1}{\Msun/\parsec}^{-3}$ (see \Cref{fig:vrot}).

\section{The S2 gravitational redshift}\label{sec:4}

The instruments on the ESO Very Large Telescope (VLT) SINFONI, NACO and more recently GRAVITY have accumulated exquisite data on the radial velocity (the redshift function) and motion of S2 for about three decades \citep{2017ApJ...837...30G,2018A&A...615L..15G}. This has allowed the recent observational {detection} of the combined gravitational redshift and relativistic transverse Doppler effect for S2 by the GRAVITY Collaboration \citep{2018A&A...615L..15G}.

The total Doppler shift $z(r)$ is a combination of the gravitational redshift and the relativistic Doppler shift. {The GRAVITY Collaboration \citep{2018A&A...615L..15G} uses the second-order post-Newtonian (2PN) expansion of the redshift function for the case of a test particle around a Schwarzschild BH. We now summarize their treatment and refer the reader to \citet{2006ApJ...639L..21Z,Do664} for its details, {while} refer to \Cref{sec:SM2} for details on the full general relativistic treatment and a derivation of the 2PN approximation. At 2PN order,  the redshift function is $z(r) \approx z_g(r) + z_D(r) + {\cal O}(1/c^2)$. The first term $z_g$ is the 2PN expression of the pure gravitational redshift} $z_g(r) = \sqrt{g_{00}(R)/g_{00}(r)} - 1\approx M_{BH}/r$, where $r$ is the position of the emitted photon (emitter/source), $R$ is the position of the receiver and $g_{00}$ is the $0$-$0$ component of the spacetime metric. {Since $R = D_\odot = 8$~kpc is the Sun distance to the Galactic center, $r\ll R$, so we have safely approximated $r/R\to 0$. The second term $z_D$ of the 2PN redshift can be split into the Keplerian (Newtonian) contribution, $z_K(r)$, and the purely relativistic transverse Doppler shift, $z_{tD}$, namely $z_D(r) \approx z_K(r) + z_{tD}(r)$. Here, $z_K(r) = \vec{v}\cdot \vec{n}$, where $\vec{n}$ is the unity vector in the direction of the line of sight, and $z_{tD}(r) = v(r)^2/2$ {(see \Cref{sec:SM2})}.
Summarizing, at 2PN order,} $z(r) = z_K(r) + z_{\rm GR}(r)$, where $z_{\rm GR}(r) = z_{tD}(r) + z_g(r)$ is the total general relativistic correction. Therefore, the deviation from a purely Newtonian behavior can be measured by the general relativistic ``excess'' of the radial velocity, $\Delta z(r) \equiv z(r) - z_K(r) = z_{\rm GR}(r)$ \citep{2018A&A...615L..15G}. {Since the extended-RAR model is fully general relativistic, we use the full general relativistic expression of the redshift function and the corresponding general relativistic excess (see \Cref{sec:SM2} for details).}

\begin{figure*}%
	\centering%
    \includegraphics[width=0.8\hsize,clip]{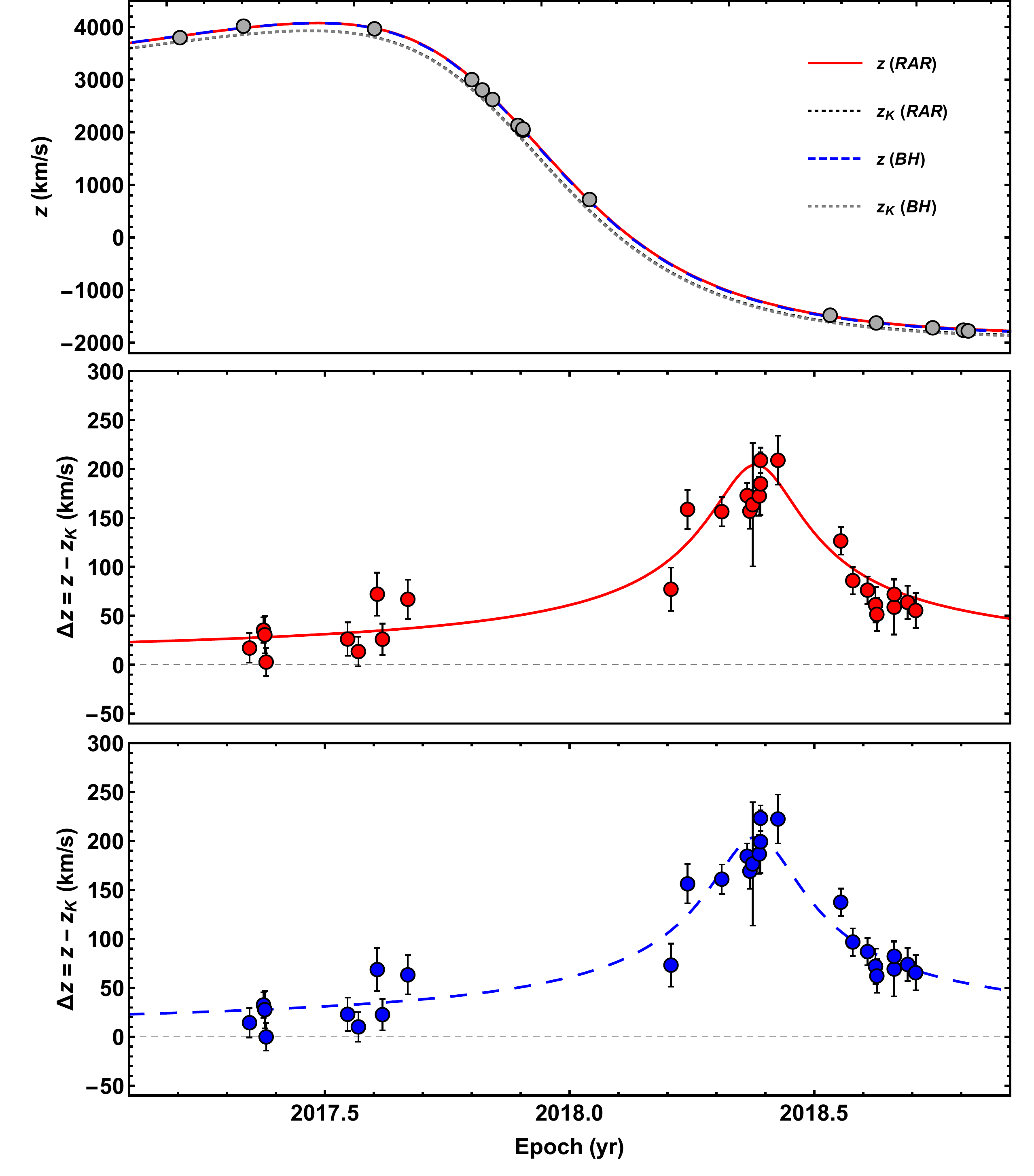}
	\caption{{Redshift function $z$ (top panel) and redshift function ``excess'' (middle panel for the RAR model and lower panel for the central massive BH model) with respect to the Keplerian (Newtonian) contribution, i.e. $\Delta z = z - z_K$ (see \Cref{sec:SM2}), for the S2 motion at around its pericenter passage. The theoretical models are calculated by solving the equations of motion of a test particle in the gravitational field of: 1) a Schwarzschild BH of \SI{4.075E6}{\Msun} (blue dashed curves), and 2) the DM distribution obtained from the extended RAR model for \SI{56}{\kilo\eV}-fermions (red curves). The mass of the quantum core in the RAR model {is \SI{3.5E6}{\Msun}}. \Cref{tab:parameter} shows the parameters of each model.}}\label{fig:RedShift}%
\end{figure*}

{
Figures \ref{fig:S2z} {and \ref{fig:G2vel} show, respectively for S2 and G2,} the redshift function $z$ computed in full general relativity, for the massive BH and the extended-RAR model. In the top panel of \Cref{fig:RedShift}, we show {for S2} in the two models, the redshift function $z$ together with the corresponding Keplerian contribution $z_K$. The bottom panels show the corresponding general relativistic excess, $\Delta z$. It can be seen from all these plots that both models fit with comparable accuracy the data. In fact, the reduced-$\chi^2$ for the redshift function for this set of parameters are: $\bar{\chi}^2_{z, \rm RAR}\approx 1.28$ and $\bar{\chi}^2_{z, \rm BH}\approx 1.04$; see \Cref{sec:SM3} for details on the calculation of $\bar{\chi}^2$. It is important to mention that there are sets of parameters, in both models, with slightly different values than the ones presented in \Cref{tab:parameter}, which produce $\bar{\chi}^2_{z,\rm RAR}\approx \bar{\chi}^2_{z,\rm BH}\approx 1$. However, those models slightly increase the $\bar{\chi}^2_X$ and $\bar{\chi}^2_Y$, so increasing the mean $\langle \bar{\chi}^2\rangle$.
}

%

\section{Discussion \& Conclusions}\label{sec:5}

The vast amount of high-precision data (position and velocity) collected in the last decade of {objects orbiting Sgr~A*, such as S2 and G2, offers} an unprecedented opportunity to test alternative scenarios to the central BH in our Galaxy. In the case of the present work, such a motivation is two-folded. First, it has been recently shown \citep{2019IJMPD..2843003A,2018PDU....21...82A,2019PDU....24100278} that fermionic DM, which self-consistently accounts for the Pauli principle and particle escape effects in the underlying phase-space DF at DM halo formation, leads to novel \textit{dense~core~--~diluted~halo} profiles where the degenerate core can produce analogous gravitational effects of a central BH. Second, the post-pericenter passage of G2 challenges the BH scenario, since in order to explain the G2 data within that picture, \citet{2019ApJ...871..126G} had to introduce an \textit{ad-hoc} drag force acting onto G2, caused by its motion through an accretion flow. In addition, for such a drag-force hypothesis to work, it is necessary that G2 be a gas cloud. Such a scenario contrasts with the observations and results of \citet{2014ApJ...796L...8W} which rule out the gas cloud composition, in favor of a stellar nature. {Moreover, even assuming G2 to be a gas cloud, and if a {radiatively inefficient} accretion flow (RIAF) is also assumed} (as done in \citealp{2019ApJ...871..126G}), the strength of the drag force onto G2 needed to explain the post-pericenter observations, implies an ambient density {$n_0 \sim$ few $10^3$~cm$^{-3}$ at $\sim 10^3 \,r_{\rm Sch}$.} However, such density value at these pericenter scales, exceeds by nearly one order of magnitude the upper bound found in recent high-resolution numerical simulations\footnote{{There are systematic uncertainties in the estimation of $n_0$ in \citet{2019ApJ...871..126G} mainly due to the unknown size of the putative gas cloud, the density profile, and the physics of the accretion process.}} \citep{2018MNRAS.473.1841S}. Such an upper bound has been obtained from the constraint that G2 be not tidally disrupted at its pericenter passage.

Turning to the core-halo DM profiles, formation scenarios in which the quantum nature of the particle is considered (i.e. either bosonic or fermionic), are still an open field of research, and our aim here is to provide a further (precision) test for fermionic models. Joint observational tests based on additional physics, e.g. strong lensing \citep{2016PhRvD..94l3004G} or DM-active neutrino interactions \citep{2020EPJC...80..183P}, can help in unambiguously probing the existence of a central fermionic DM concentration in the allowed region of the extended RAR model parameter space. The results shown here imply that such free parameter space is slightly reduced {with respect to} the former one given in \citet{2018PDU....21...82A}. For fermion masses below \SI{56}{\kilo\eV}, the size of the DM core increases and there is also orbital precession. Thus, data of the orbital precession of S2 {\citep{2020A&A...636L...5G} might further {constrain} the allowed range of the fermion mass}. The other free parameters are well within the allowed range as broadly constrained in \citet{2019PDU....24100278} for each galaxy type.

In this work, we have {used} the existing observational data of S2 including the total Doppler shift, which has both special and general relativistic contributions, and the orbit in the plane of sky and its radial velocity. We have solved the equations of motion for a test particle ({S2 and G2}) in the gravitational field {produced by} two cases of interest: 1) the central massive BH hypothesis for which we have used the Schwarzschild metric, and 2) {the fermionic DM hypothesis} within the extended-RAR model, which leads to a DM core-halo profile leading to a metric obtained from the extended RAR model equilibrium equations following the treatment in \citet{2018PDU....21...82A} and summarized in \Cref{sec:SM1}. We refer to \Cref{sec:SM3} for details on the equations of motion and the procedure to obtain the model parameters from the fitting of the observational data.
{
We have found that in the case of S2, both the massive BH model and the RAR model can explain all the observational data (orbit and velocity) with comparable accuracy, but the RAR model is preferable with a lower $\langle \bar{\chi}^2 \rangle$; see \Cref{tab:parameter}, \Cref{fig:S2} and \Cref{fig:S2z}, including the general relativistic redshift, see \Cref{fig:RedShift}. In the case of G2, only the RAR model can explain both the orbit and velocity, see  \Cref{tab:parameter}, \Cref{fig:G2} and \Cref{fig:G2vel}.
}

This remarkable result {of the extended-RAR fermion-DM model} is further complemented with the successful applicability of its ensuing \textit{dense~core~--~diluted~halo} profile to other galaxy types, from dwarfs to ellipticals \citep{2019PDU....24100278}. Moreover, it can be directly linked with the DM-halo formation processes since, the RAR model quantum-statistical phase-space distribution (see \Cref{fcDF} in \Cref{sec:SM1}), is not given \emph{ad-hoc} but it can be obtained as a (quasi) stationary solution of a generalized thermodynamic Fokker-Planck equation for fermions \citep{2004PhyA..332...89C}. This includes the physics of collisionless (violent) relaxation and evaporation, appropriate for non-linear structure formation. Such phase-space distributions have been there shown to fulfill a maximization (coarse-grained) entropy principle (second law of thermodynamics) during the (collisionless) relaxation process until the halo reaches the currently observed steady state. 

The present results give a strong observational support to the  \textit{quantum-core} {hypothesis} in alternative to the massive BH one in Sgr A* \citep{2019IJMPD..2843003A,2018PDU....21...82A}, and also to the {fermionic} nature of DM. In this line, besides the dynamical constraints, it is desirable to further test the presence of fermionic DM concentrations in our galactic core from existing luminosity constraints on the variability of the compact radio source Sgr A*. Such a study goes beyond the scope of the present work that is devoted to the orbital dynamics of some of the closest objects to Sgr A* and with accurate astrometric data. {We would like to recall, however, that the gravitational potentials produced by a BH and by a most compact (stable) DM quantum core for fermion mass of the order of \SI{100}{\kilo\eV}, practically coincide at distances $r\gtrsim 10 r_{\rm Sch}$ \citep[see][for details]{2016PhRvD..94l3004G}. {The dynamics of baryonic matter and its emission associated with its motion at those scales is thus not expected to differ much between the two pictures.} Differences might occur in the innermost regions owing to the `transparency' of the DM core, leading to differences in the lensing properties \citep{2016PhRvD..94l3004G}, and possibly on any accretion process at these small scales. Moreover, although the emission around Sgr~A* is often univocally associated with a particular accretion flow (extremely under luminous when compared to typical accretion expectations), this is not confirmed by the observational data, and indeed, alternative mechanisms/explanations for the observed radiation exist \citep[see, e.g.,][for a review on this subject]{2014ARA&A..52..529Y}. In fact, as of today, the most reliable observational data that allow to prove and test the validity of alternative models for Sgr~A*, as the one presented in this work, are} the precision measurements of the orbital dynamics, together with the validity and demonstrated precision of general relativity. We look forward to the public release of the latest data by the GRAVITY Collaboration, both on S2 and G2 \citep[e.g.][]{2019ApJ...871..126G}, which will serve to further test our theoretical prediction (e.g.~\Cref{fig:G2vel}). {We have shown in this work the results for a fermion mass of $\SI{56}{\kilo\eV}$, a value safely larger than the lower limit of $\SI{48}{\kilo\eV}$ estimated in \citet{2018PDU....21...82A} by equating the DM core radius to the up-to-then reported pericenter distance of S2.} The lower the fermion mass, the larger the size of the DM core, and vice-versa.  Therefore, it is worth to explore whether the data of S2 and G2 together might further constrain the allowed range of fermion masses. Such an investigation, however, goes beyond the scope of the present work, and could be a topic of joint collaboration.

The {DM-fermion mass of \SI{56}{\kilo\eV} inferred in this work} would produce (down to Mpc scales) the same standard $\Lambda$CDM power-spectrum, hence providing the expected large-scale structure \citep{2009ARNPS..59..191B}. Since the {fermion} mass is larger than {$>\SI{5}{\kilo\eV}$}, it is not in tension with constraints from the Lyman-$\alpha$ forest {\citep{2009PhRvL.102t1304B,2013PhRvD..88d3502V,PhysRevD.96.023522}} and the number of Milky Way satellites \citep{2008ApJ...688..277T}. Furthermore, for the present fermion mass $mc^2=\SI{56}{\kilo\eV}$, the critical mass for gravitational collapse of the DM quantum core ($M_c^{\rm cr}\sim  m_{\rm Pl}^3/m^2$, with $m_{\rm Pl}$ the Planck mass) into a BH is of the order of $\SI{E8}{\Msun}$, providing a viable formation scenario for the observed central supermassive BH in active galaxies such as M87. Indeed, a supermassive BH of $\sim \SI{E9}{\Msun}$ can form starting from a $\sim \SI{E8}{\Msun}$ BH-seed and accreting $\lesssim 1 \%$ of the (baryonic and/or DM) galactic environment of $\sim \SI{E12}{\Msun}$. Over cosmological timescales, this would be achieved without unrealistic super-Eddington accretion rates, {while providing a new framework to study the poorly understood formation and growth scenarios of supermassive BH seeds in the cosmological high-redshift Universe}.\\


\noindent\textit{Acknowledgments}. {We thank the Referee and the Editor for their helpful suggestions which have improved the presentation of our results.} E.A.B-V. thanks financial and research support from COLCIENCIAS under the program Becas Doctorados Nacionales 727, the International Center for Relativistic Astrophysics Network (ICRANet), Universidad Industrial de Santander (UIS) and the International Relativistic Astrophysics Ph.D Program (IRAP-PhD). C.R.A has been supported by CONICET and Secretary of Science and Technology of FCAG.



\begin{appendix} 

\section{The extended Ruffini-Arg\"uelles-Rueda (RAR) model}\label{sec:SM1}

The extended-RAR model conceives the DM in galaxies as a general relativistic, self-gravitating system of massive fermions (spin $1/2$) in hydrostatic and thermodynamic equilibrium. {It uses an} equation of state (EOS) that takes into account (i) relativistic effects of the fermionic constituents, (ii) finite temperature effects and (iii) particle escape effects at large momentum ($p$) via a cut-off in the Fermi-Dirac distribution $f_c$: 
\begin{equation}
f_c(\epsilon\leq\epsilon_c) = \frac{1-e^{(\epsilon-\epsilon_c)/kT}}{e^{(\epsilon-\mu)/kT}+1}, \qquad f_c(\epsilon>\epsilon_c)=
0\, ,
\label{fcDF}
\end{equation}
{differentiating from the original RAR model version (see \Cref{sec:2}) only in the condition (iii).} Where $\epsilon=\sqrt{c^2 p^2+m^2 c^4}-mc^2$ is the particle kinetic energy, $\mu$ is the chemical potential with the particle rest-energy subtracted off, $T$ is the temperature, $k$ is the Boltzmann constant, $c$ is the speed of light, and $m$ is the fermion mass. The stress-energy tensor is the one of a perfect fluid with the density and pressure associated with this distribution function, i.e.:
\begin{align}
\label{rhoepdefscutoff}
	\rho &= m\frac{2}{h^3}\int_{0}^{\epsilon_c}f_c(p)\left(1+\frac{\epsilon(p)}{mc^2}\right)d^3p\ ,\\
	P	 &= \frac{1}{3}\frac{4}{h^3}\int_{0}^{\epsilon_c}f_c(p)\,\epsilon\,\frac{1+\epsilon(p)/2mc^2}{1+\epsilon(p)/mc^2}d^3p.
\end{align}

For the spherically symmetric spacetime metric
\begin{equation}
	\label{eqn:metric}
	ds^2 = g_{00}(r){\rm d}t^2 - g_{11}(r){\rm d}r^2 - r^2\left({\rm d}\theta^2 + \sin^2{\theta} {\rm d}\phi^2\right),
\end{equation}
where ($r$,$\theta$,$\phi$) are the spherical coordinates. Using $g_{00}(r)=e^{\nu(r)}$, the Tolman \citep{1930PhRv...35..904T}, Klein \citep{1949RvMP...21..531K}, and the cutoff \citep{1989A&A...221....4M} conditions of thermodynamic equilibrium and energy conservation are: 
\begin{align}
e^{\nu/2} T &= {\rm constant},\label{eq:cond1}\\
e^{\nu/2}(\mu+m c^2) &= {\rm constant},\label{eq:cond2}\\
e^{\nu/2}(\epsilon+m c^2) &=  {\rm constant}.\label{eq:cond3}
\end{align}

The Einstein equations together with the conditions given {by \Cref{eq:cond1,eq:cond2,eq:cond3} form} a coupled system of integro-differential equations:
\begin{align}
	\frac{d\hat M}{d\hat r}&=4\pi\hat r^2\hat\rho, \label{eq:eqs1}\\
	\frac{d\theta}{d\hat r}&=-\frac{1-\beta_0(\theta-\theta_0)}{\beta_0}
    \frac{\hat M+4\pi\hat P\hat r^3}{\hat r^2(1-2\hat M/\hat r)},\label{eq:eqs2}\\
    \frac{d\nu}{d\hat r}&=\frac{2(\hat M+4\pi\hat P\hat r^3)}{\hat r^2(1-2\hat M/\hat r)}, \\
    \beta(\hat r)&=\beta_0 e^{\frac{\nu_0-\nu(\hat r)}{2}}, \\
    W(\hat r)&=W_0+\theta(\hat r)-\theta_0\, ,\label{eq:Cutoff}
\end{align}
where the subscript `$0$' stands for variable evaluated at $r=0$, and we have introduced dimensionless quantities: $\beta=k T/(m c^2)$, $\theta=\mu/(k T)$, $W=\epsilon_c/(k T)$, $\hat r=r/\chi$, $\hat M=G M/(c^2\chi)$, $\hat\rho=G \chi^2\rho/c^2$, $\hat P=G \chi^2 P/c^4$, where $\chi=2\pi^{3/2}(\hbar/mc)(m_{\rm Pl}/m)$ being $m_{\rm Pl}=\sqrt{\hbar c/G}$ the Planck mass.

This system is solved for appropriate boundary conditions, $[M(0)=0,\theta(0)=\theta_0,\beta(0)=\beta_0,\nu(0)=0,W(0)=W_0]$, for different DM particle masses $m$, to find a solution consistent with the DM halo observables of a given galaxy. The RAR models equations are solved for positive central degeneracy parameters (i.e. $\theta_0 > 10$) in order to {ensure that} the Pauli principle is fulfilled within the central core, as demonstrated in \citep{2015MNRAS.451..622R,2018PDU....21...82A}. This property implies as a consequence RAR DM profiles which develop a \textit{dense~core~--~diluted~halo} morphology, where the central core is governed by Fermi-degeneracy pressure, while the outer halo holds against gravity by thermal pressure (resembling the Burkert or King profiles as shown in \citealp{2018PDU....21...82A,2019PDU....24100278}). Indeed the extended RAR model is the more general of its kind, given it does not work under the fully-Fermi-degeneracy approximation as in \citep{2017MNRAS.467.1515R}, nor in the diluted-Fermi regime \citep{2014MNRAS.442.2717D}.

The case of the Milky Way has been recently analyzed in \citet{2018PDU....21...82A}. {We adopt here a similar boundary condition problem as solved in \citet{2018PDU....21...82A}, with the only difference that we now allow for the dense DM core $M_c$ to vary until the mean reduced-$\chi^2$ of the S2 data fit (see \Cref{sec:SM3}) achieve the minimum. That is, we consider}: (i) a DM halo mass with {observationally inferred values} at two different radial locations in the Galaxy: a DM halo mass $M(r=40~{\rm kpc}) = \SI{2E11}{\Msun}$ \citep{2014MNRAS.445.3788G} and $M(r=12~{\rm kpc}) = \SI{5E10}{\Msun}$ \citep{2013PASJ...65..118S}; {and (ii) a DM dense quantum core to have a mass $M(r=r_c)\equiv M_c= \SI{3.5E6}{\Msun}$ with $r_c$ smaller than the S2 star pericenter, resulting in $r_c\approx 0.4$~mpc by the extended-RAR model free parameters given in \Cref{fig:vrot}. While the halo condition (i) follows exactly the methodology used in \citet{2018PDU....21...82A}, the latter condition (ii) explicitly request the quantum DM core to substitute the massive BH scenario, while minimizing the mean reduced-$\chi^2$ for the S2 data fit (see \Cref{sec:SM3})}. We have thus three boundary conditions for three free RAR-model parameters ($\beta_0$, $\theta_0$, $W_0$), for a given particle mass {of $m c^2=\SI{56}{\kilo\eV}$. It is of interest to explore whether the data of S2 and G2 together can further constrain the allowed range of fermion masses. Such an investigation, however, goes beyond the scope of the present work}. The application {of the extended-RAR model} to other galaxy types from dwarfs to ellipticals to galaxy clusters can be found in \citet{2019PDU....24100278}.

\section{Total orbital Doppler shift}\label{sec:SM2}

{
The redshift is defined by the ratio between the measured wavelength of a spectral line at emission and reception:
\begin{equation}\label{eq:z}
    1 + z \equiv \frac{{\cal E}_{\rm (em)}}{{\cal E}_{\rm (obs)}} = \frac{\lambda_{\rm (obs)}}{\lambda_{\rm (em)}}.
\end{equation}
We denote the four-momentum of photons measured by an observer comoving with the emitter, $k^\mu_{\rm (em)}$, and the one measured by an observer comoving with the receiver, $k^\mu_{\rm (obs)}$. The observer comoving with the emitter has four-velocity $u^\mu_{\rm (em)}$, {so they measure a photon energy} ${\cal E}_{\rm (em)} = k_\mu^{\rm (em)}u^\mu_{\rm (em)}$. Analogously, the observer comoving with the receiver measures a photon energy ${\cal E}_{\rm (obs)} = k_\mu^{\rm (obs)}u^\mu_{\rm (obs)}$. Therefore, theoretically, we can write \Cref{eq:z} as:
\begin{equation}\label{eqn:zGR}
1 + z = \frac{k_\mu^{\rm (em)} u^\mu_{\rm (em)}}{k_\mu^{\rm (obs)} u^\mu_{\rm (obs)}} = \frac{k^{\rm (em)}_0}{k^{\rm (obs)}_0}\frac{u^0_{\rm (em)} + u^i_{\rm (em)} n^{\rm (em)}_i}{u^0_{\rm (obs)} + u^i_{\rm (obs)} n^{\rm (obs)}_i},
\end{equation}
where $n^i = k^i/k^0$ are the normalized spatial components of the photon four-momentum. Defining the components of the three-velocity, $v^i \equiv u^i/u^0$, and the Lorentz factor {(where the right-hand side of the equation below is obtained from the normalization condition $u^\mu u_\mu = 1$)}:
\begin{equation}\label{eqn:gamma}
    \gamma =  u^0 = \frac{d t}{d\tau} = \frac{1}{\sqrt{g_{00} - v^2}},\quad v^2 = -v^i v_i = -g_{11} (v_r)^2 + (r v_\phi)^2,
\end{equation}
{then} \Cref{eqn:zGR} becomes
\begin{equation}\label{eqn:z1}
1 + z = \frac{\gamma_{\rm (em)}}{\gamma_{\rm (obs)}}\frac{1 + v^i_{\rm (em)} n^{\rm (em)}_i}{1 + v^i_{\rm (obs)} n^{\rm (obs)}_i},
\end{equation}
where we have used the fact that along the photon geodesic $k_0$ is conserved.
}

{
For the present purpose, with sufficient accuracy, one can neglect the motion of the observer reference frame with respect to the one of the center of the gravitational field, i.e. $v^i_{\rm (obs)} = 0$, and the gravitational field at the observation point, $g^{\rm (obs)}_{00} = 1$ \citep[see, e.g.,][]{Do664}, then $\gamma_{\rm (obs)} = 1$ and \Cref{eqn:z1} becomes:
\begin{equation}\label{eqn:zfinal}
1 + z = \gamma\,(1 + \vec{v}\cdot \vec{n}),
\end{equation}
where $\vec{v}\cdot \vec{n} = v^i n_i$ is the three-dimensional velocity of the emitter projected onto the direction of the line of sight, i.e. what is often called in the experimental literature as the observed ``radial velocity'', and we have relaxed the notation of emitter and receiver since only the emitter is being considered in motion.
}

{It is important to clarify that the redshift function $z$ is often referred in the literature to as ``radial velocity'', the velocity in the direction of the {line of sight}. The latter is actually $\vec{v}\cdot \vec{n}$, so as it can be seen from \Cref{eqn:zfinal}, the relation between it and $z$ is in general non-linear.}

{
In general, it is not possible to separate the contributions to $z$ of the gravitational field and of the emitter/receiver relative motion, i.e. they are combined/mixed in \Cref{eqn:zfinal}. However, this equation shows already explicitly that, in the non-relativistic limit ($\gamma \to 1$), the redshift is given only by the so-called Keplerian (Newtonian) contribution, i.e. $z \to z_K$ where:
\begin{equation}\label{eqn:zK}
z_K \equiv \vec{v}\cdot\vec{n}.
\end{equation}
}

{
The gravitational and relative motion contributions clearly show up when performing a post-Newtonian expansion of the redshift. For instance, in the case when the gravitational field is produced by a Schwarzschild BH of mass $M_{\rm BH}$, i.e. $g_{00} = -1/g_{11} = 1 - 2M_{\rm BH}/r$, the Lorentz factor, up to order $1/c^2$ (i.e. 2PN order), is:
\begin{align}\label{eqn:gamma2PN}
    \gamma &\approx \left(1 +\frac{M_{\rm BH}}{r}\right)\left(1 + \frac{v^2/2}{1-2 M_{\rm BH}/r}\right)\nonumber\\
    &\approx \left(1 +\frac{M_{\rm BH}}{r}\right)\left[1 +\frac{1}{2}v^2 \left(1 + \frac{2 M_{\rm BH}}{r}\right)\right]\nonumber \\
    &\approx \left(1 +\frac{M_{\rm BH}}{r}\right)\left(1 +\frac{1}{2}v^2\right)\approx 1 + \frac{1}{2}v^2 +\frac{M_{\rm BH}}{r} + \mathcal{O}(1/c^2),
\end{align}
which replaced into \Cref{eqn:zfinal} leads to the 2PN redshift function:
\begin{equation}\label{eqn:zPPN}
    z \approx z_K + \frac{1}{2}v^2 + \frac{M_{\rm BH}}{r} + \mathcal{O}(1/c^2).
\end{equation}
\Cref{eqn:zPPN} is the expression presented in \citet{2006ApJ...639L..21Z} (see Eq. 1 therein), and it is the radial velocity equation (S24) in \citet{Do664}, setting $v_{z0}=0$ there, and consistent with our assumption of neglecting the relative motion of the gravitational center of mass with respect to the center of the observer’s reference frame. The approximate \Cref{eqn:zPPN} has been used in those works for the analysis of the gravitational contribution to the redshift function in the case of the S2 star.
}

{
The GRAVITY Collaboration \citep{2018A&A...615L..15G}, has claimed the detection of the gravitational redshift in the orbit of the star S2. In practice, they verify the consistency of the data of the redshift function of S2 with the presence of what they call the ``general relativistic excess of the radial velocity'' \citep{2018A&A...615L..15G}:
\begin{equation}\label{eq:Deltaz}
    \Delta z \equiv z - z_K,
\end{equation}
\Cref{eq:Deltaz} tells that the theoretical excess predicted by general relativity at 2PN order is:
\begin{equation}\label{eq:Deltaz2PN}
    \Delta z \approx \frac{1}{2}v^2 + \frac{M_{\rm BH}}{r}, 
\end{equation}
which has been shown to be consistent with the data of the S2 star \citep{2018A&A...615L..15G}.
}

{
The present RAR model is a fully general relativistic treatment, therefore we use the full redshift function (\ref{eqn:zfinal}) in the fit of the observational data (see \Cref{sec:SM3}). In this case, the general relativistic excess in the redshift, as defined by \Cref{eq:Deltaz}, reads:
\begin{equation}\label{eq:DeltazRAR}
    \Delta z = (\gamma-1)(1 + z_K).
\end{equation}
It is manifest in the fully general expression \Cref{eq:DeltazRAR} that in the non-relativistic (Newtonian) limit, $\gamma \to 1$, the excess vanishes, i.e. $\Delta z \to 0$. It is also easy to check that \Cref{eq:DeltazRAR} reduces to \Cref{eq:Deltaz2PN} at 2PN order, with the aid of \Cref{eqn:gamma2PN}.
}

\section{Equations of motion {and orbital parameters of the real and apparent orbits}}\label{sec:SM3}
 
\subsection{Orbital dynamics} 
 
The equations of motion of the test particle (S2 or G2), in the spherically symmetric metric given by \Cref{eqn:metric}, assuming without loss of generality $\theta= \pi/2$, are: 
\begin{subequations}
    \begin{eqnarray}
        \dot{t} &=& \dfrac{E}{g_{00}(r)},\label{eqn:motiont}\\
        \ddot{r} &=& \dfrac{1}{2 \  g_{11}(r)}\left[\dfrac{d g_{00}(r)}{dr} \ \dot{t}^2 - \dfrac{d g_{11}(r)}{dr} \ \dot{r}^2 - 2 \ r \ \dot{\phi}^2\right],\label{eqn:motionr}\\
        \dot{\phi} &=&  \dfrac{L}{r^2},\label{eqn:motionphi}
\end{eqnarray}
\end{subequations}
where $E$ and $L$ are the conserved energy and the angular momentum of the particle per-unit-mass, {so $E$ is dimensionless and $L$ has units of mass}, and the overdot stands for derivative with respect to the proper time, $\tau$. 
{
In terms of Cartesian coordinates, we denote the position and velocity components of the real orbit as: $x$, $y$, $z$, and $v_x$, $v_y$, $v_z$. In our present case, $\theta = \pi/2$, these are obtained using the transformation from spherical Schwarzschild coordinates to Cartesian coordinates:
\begin{align}\label{eqn:xyz}
    x &= r \cos\phi,\\
    y &= r \sin\phi,\\
    z &= 0,
\end{align}
and the corresponding three-velocities are:
\begin{align}\label{eqn:vxvyvz}
     v_x &= v_r \cos\phi - r v_\phi \sin\phi,\\
     v_y &= v_r \sin\phi + r v_\phi \cos\phi,\\
     v_z &= 0,
\end{align}
where $v_r \equiv u^r/u^0 = dr/dt$ and $v_\phi\equiv u^\phi/u^0 = d\phi/dt$, being $u^\mu = d x^\mu/d\tau$ the particle's four-velocity.
}

\begin{figure}%
	\centering%
	\includegraphics[width=\hsize,clip]{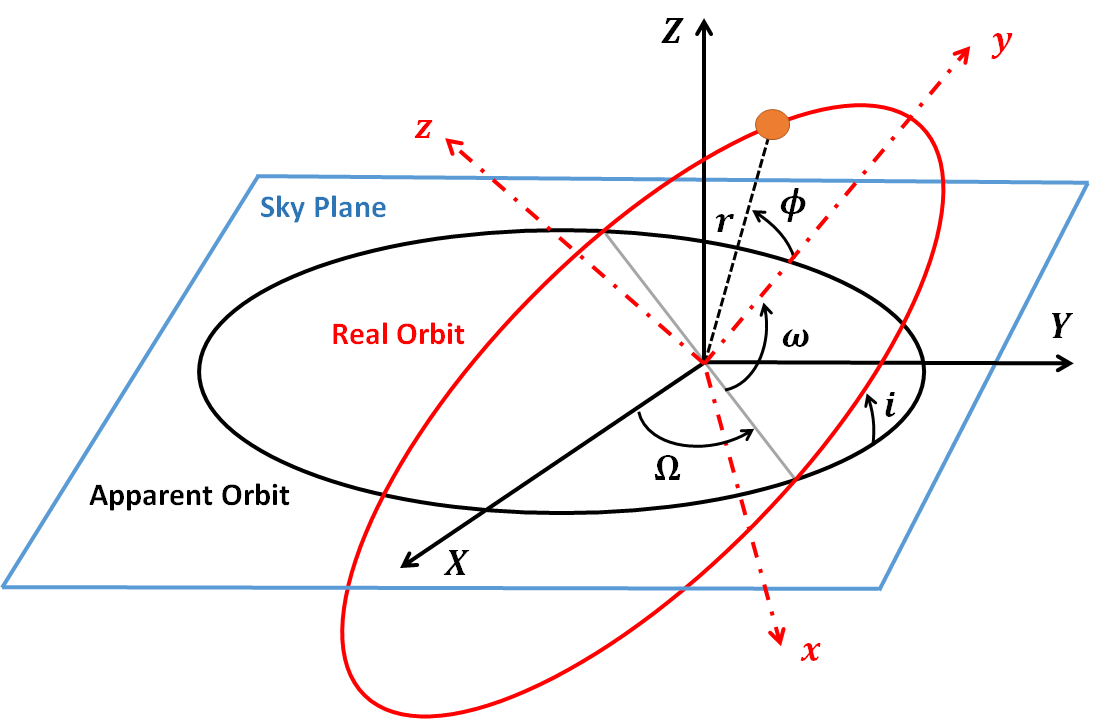}
	\caption{Projection of real orbit onto the plane of the sky. The axes originate at Sgr A* (the focus of the ellipse). The picture shows an illustration of the orbital parameters: $\phi$ {is the azimuth angle of the spherical system of coordinates associated with the $x$, $y$, $z$ Cartesian coordinates, i.e. for an elliptic motion in the $x$-$y$ plane, it is the true anomaly}, $i$ is the angle of inclination between the real orbit and the observation plane, $\Omega$ is the angle of the ascending node and $\omega$ is the argument of pericenter. It is worth noting that the $Z$-axis of the coordinate system is defined by the vector pointing from the solar system to the galactic center.}\label{fig:orbital_parameters}%
\end{figure}

The solution of eqs.~(\ref{eqn:motiont})--(\ref{eqn:motionphi}) allows to trace the stellar orbit, however, to compare with the observational data, it is necessary to determine the apparent orbit on the plane of the sky. Namely, we have to project the real orbit onto the observation plane as shown in \Cref{fig:orbital_parameters}. {On the plane of the sky, the star traces an orbit with Cartesian positions $X_{\rm obs}$ and $Y_{\rm obs}$, defined by the observed angular positions, i.e. the declination $\delta$ and the right ascension $\alpha$ \citep[see, e.g.][]{2008ApJ...689.1044G, 2018ApJ...854...12C, Do664}:
\begin{equation}\label{eqn:XY}
    X_{\rm obs} = D_\odot (\alpha-\alpha_{\rm Sgr A*}),\quad Y_{\rm obs} = D_\odot (\delta-\delta_{\rm Sgr A*})
\end{equation}
centering the coordinate system on Sgr A*. We adopt in this work $D_\odot = 8$~kpc \citep[see, e.g.,][]{2018A&A...615L..15G,Do664}.
}

Introducing the same notation of \citet{Do664} for the classic Thiele-Innes constants, i.e. $A$, $B$, $C$, $F$, $G$, $H$, the theoretical apparent orbit (i.e. the position in coordinates $X$, $Y$, $Z$), can be obtained from the real orbit positions $x$ and $y$, by (see \Cref{fig:orbital_parameters}):
\begin{subequations}\label{eqn:XYZ}
\begin{align}
    X &=  x\,B + y\, G,\label{eqn:XX}\\
    Y &= x\, A + y\, F,\label{eqn:YY}\\
    Z &= x\, C + y\, H,\label{eqn:ZZ}
\end{align}
\end{subequations}
and the corresponding components of the apparent coordinate velocity are:
\begin{subequations}\label{eqn:vXvYvZ}
\begin{align}
    V_X &= \frac{d X}{dt} = v_x B + v_y G,\\
    V_Y &= \frac{d Y}{dt} = v_x A + v_y F,\\
    V_Z &= \frac{d Z}{dt} = v_x C + v_y H,
\end{align}
\end{subequations}
where
\begin{subequations}\label{eqn:constants}
    \begin{eqnarray}
    A &=& \cos\Omega \cos\omega - \sin\Omega \sin\omega \cos i,\\
    B &=& \sin\Omega \cos\omega + \cos\Omega \sin\omega \cos i,\\
    C &=& \sin\omega \sin i,\\
    F &=& -\cos\Omega \sin\omega - \sin\Omega \cos\omega \cos i,\\
    G &=& -\sin\Omega \sin\omega + \cos\Omega \cos\omega \cos i,\\
    H &=& \cos\omega \sin i,
    \end{eqnarray}
\end{subequations}
{being $\omega$, $i$, and $\Omega$ the \emph{osculating} orbital elements, respectively the argument of pericenter, the inclination between the real orbit and the observation plane, and the ascending node angle. These orbital elements are strictly defined (fixed constants) only for} a Keplerian (Newtonian) elliptic orbit. {In that case,} the radial position is simply given by $r = a (1-e \cos E)$, where $a$ is the semi-major axis of the ellipse, $e$ its eccentricity, and $E$ its eccentric anomaly. The latter is related to the true anomaly, which is the azimuthal angle $\phi$, by $\cos\phi = (\cos E - e)/(1-e\cos E)$. In such a case, \Cref{eqn:XYZ} and \Cref{eqn:vXvYvZ} reduce to the eqs.~(S8)--(S10) of \citet{Do664}. However, in the full general relativistic case, it is not possible to find (in general) a closed-form with an analytic function $r(\phi)$ describing the orbit. For the simpler case of a test-particle moving around a Schwarzschild BH, $r(\phi)$ can be written in terms of Jacobi elliptic functions. In the case of the RAR model, we obtain $r(\tau)$ and $\phi(\tau)$ or, for the sake of comparison with observations, $r(t)$ and $\phi(t)$, by numerical integration of the equations of motion, \Cref{eqn:motiont}. Clearly, we can then obtain $r(\phi)$ numerically.

\subsection{Fitting procedure of the observational data}

{
For the fitting of the observed positions, \citet{Do664} introduce time-varying offsets of the position of the gravitational center of mass with respect to the center of the reference frame, adopting a linear drift. For our purpose, it is sufficient to introduce the constant offsets $X_0$ and $Y_0$, i.e.:
\begin{subequations}
\begin{align}
    X_{\rm obs}(t_{\rm obs}) &= X[r(t), \phi(t);\omega,i, \Omega] + X_0,\label{eqn:XYobsXY_X0}\\
    Y_{\rm obs}(t_{\rm obs}) &= Y[r(t), \phi(t);\omega,i, \Omega]+ Y_0,\label{eqn:XYobsXY_Y0}
\end{align}
\end{subequations}
where $X$ and $Y$ are given by \Cref{eqn:XYZ}, $t_{\rm obs}$ is the time measured at the {observer} point, and $t=t_{\rm em}$ is the time at emission.
}

\begin{figure}
    \centering
    \includegraphics[width=\hsize,clip]{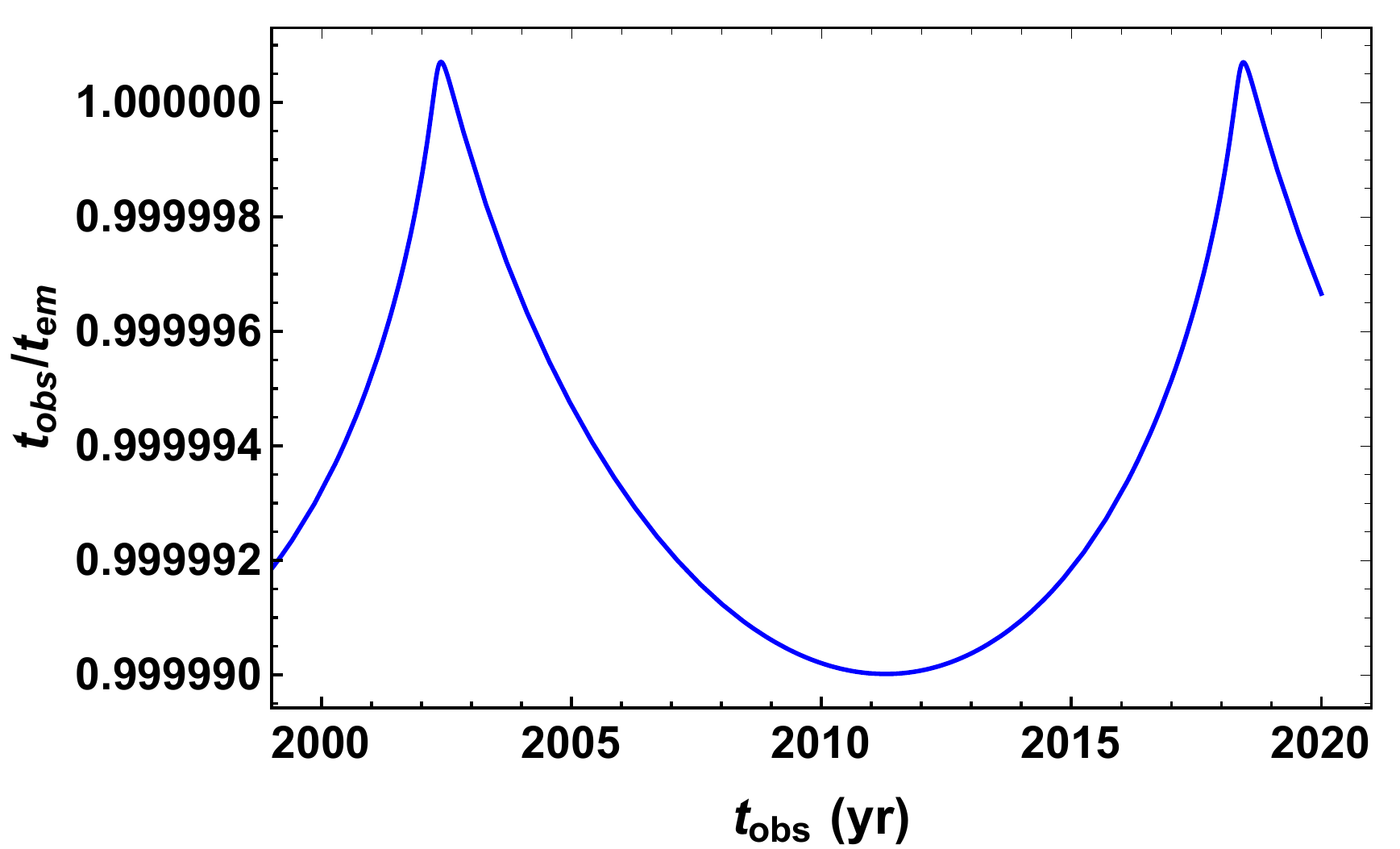}
    \caption{{Ratio $t_{\rm obs}/t_{\rm em}$ as given by \Cref{eqn:tem}, calculating $Z(t_{\rm obs})$ with eq.~(\ref{eqn:ZZ}), for the best-fit model parameters of the BH model \Cref{tab:parameter}, derived assuming \Cref{eqn:temtobs}.}}
    \label{fig:temtobs}
\end{figure}

{
In general, $t_{\rm obs}$ and $t$ are not equal, namely there exist a time delay in the observations due to light-propagation effects along the line of sight. An obvious cause of time delay is the fact that the speed of light is finite. Along the {line-of-sight} direction (i.e. the $Z$-direction), this is called Rømer delay \citep[see e.g.][]{Do664}:
\begin{equation}\label{eqn:tem1}
    t_{\rm obs} = t_{\rm em} + \frac{Z(t_{\rm em})}{c},
\end{equation}
where $Z$ is given by eq.~(\ref{eqn:ZZ}). The \Cref{eqn:tem1} is an implicit non-linear equation for $t_{\rm em}$ but it can be inverted at first order as \citep[see e.g.][]{Do664}:
\begin{equation}\label{eqn:tem}
    t_{\rm em} \approx t_{\rm obs} - \frac{Z(t_{\rm obs})}{c}.
\end{equation}
In our fitting procedure, we neglect any photon delay time, so we adopt:
\begin{equation}\label{eqn:temtobs}
    t_{\rm em} = t_{\rm obs},
\end{equation}
which is sufficiently accurate for the purposes of the present work. Indeed, the model parameters we have inferred (see \Cref{tab:parameter}) of S2 in the case of a Schwarzschild BH, are similar to the ones previously presented in the literature; see e.g. \citet{2018A&A...615L..15G} and \citet{Do664} for comparison. In fact, as shown in \Cref{fig:temtobs}, $t_{\rm obs}\approx t_{\rm em}$ with high accuracy ($\approx 0.001\%$ error). {Our estimate shown in \Cref{fig:temtobs} agrees with the one in \citet{Do664}, who mentioned this delay modulates the light-propagation time by $\Delta t = t_{\rm obs}-t_{\rm em} \approx -0.5$~days at pericenter and $\Delta t \approx  7.5$~days at apocenter}.
}

{
The assumption of zero relative motion of the center of mass and the center of the observer's frame, introduces only a difference of order $v_{z0}/v_Z \sim 0.1\%$ in the radial velocity, being $z$ the redshift function (see \Cref{sec:SM2} for details).
}

{
In general, the four-velocity component $u_Z$ is not directly accessible from the observations, as it is the redshift function $z$ given by \Cref{eqn:zfinal}. Therefore, we obtain the parameters that best fit the equation
\begin{equation}\label{eqn:zobs}
    z_{\rm obs}(t_{\rm obs}) = z[r(t), \phi(t), \dot{r}(t), \dot{\phi}(t);\omega,i],
\end{equation}
where, in terms of the orbital parameters:
\begin{subequations}\label{eqn:zuz}
\begin{align}
    z &= \gamma -1 + u_Z, \\
    u_Z &= \gamma V_Z = \left[\dot{r} \  \sin(\phi+\omega) + r \dot{\phi} \ \cos(\phi + \omega)\right] \sin i.
\end{align}
\end{subequations}
in which we have introduced the notation $V_Z \equiv \vec{v}\cdot \vec{n}$, being $\vec{n}$ the unit vector pointing from the emitter to the observer (i.e unit vector in the direction of the line of sight), and we recall that $\dot{r} = dr/d\tau$, $\dot{\phi} = d\phi/d\tau$, and $\gamma$ is given by \Cref{eqn:gamma}.
}

{
It is now clear that, at every time, the possible available observational data are: the coordinates of the apparent orbit in the sky plane, i.e. $X_{\rm obs}$ and $Y_{\rm obs}$, and the measured redshift function, $z_{\rm obs}$. The real orbit, at every time, is obtained by solving the equations of motion,  \Cref{eqn:motiont}, which give the coordinate positions $r(t)$, $\phi(t)$, and the corresponding velocities $\dot{r}(t)$ and $
\dot{\phi}(t)$.
}

\begin{figure*}[htbp!]
    \centering
    \includegraphics[width=0.49\hsize,clip]{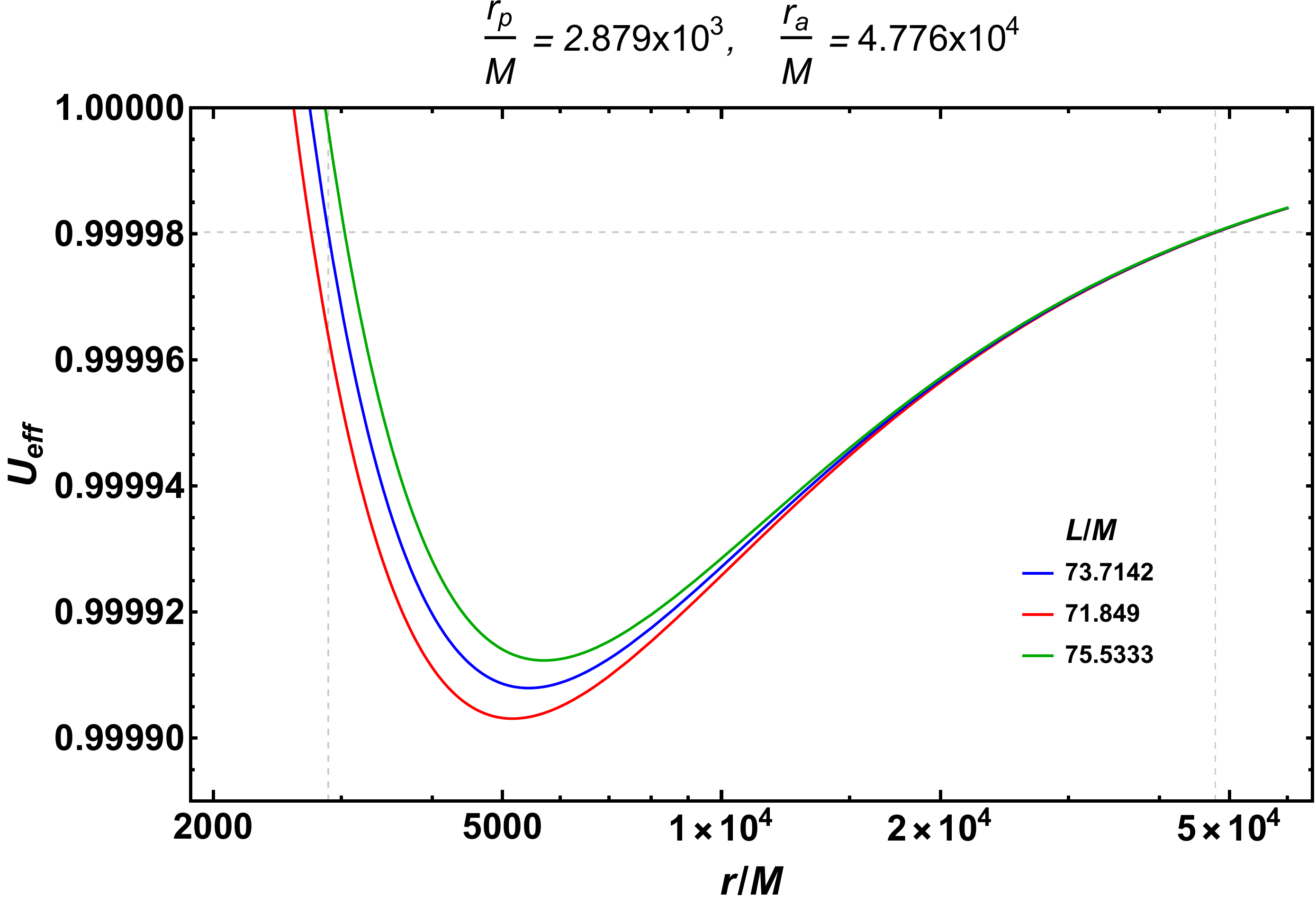}\includegraphics[width=0.49\hsize,clip]{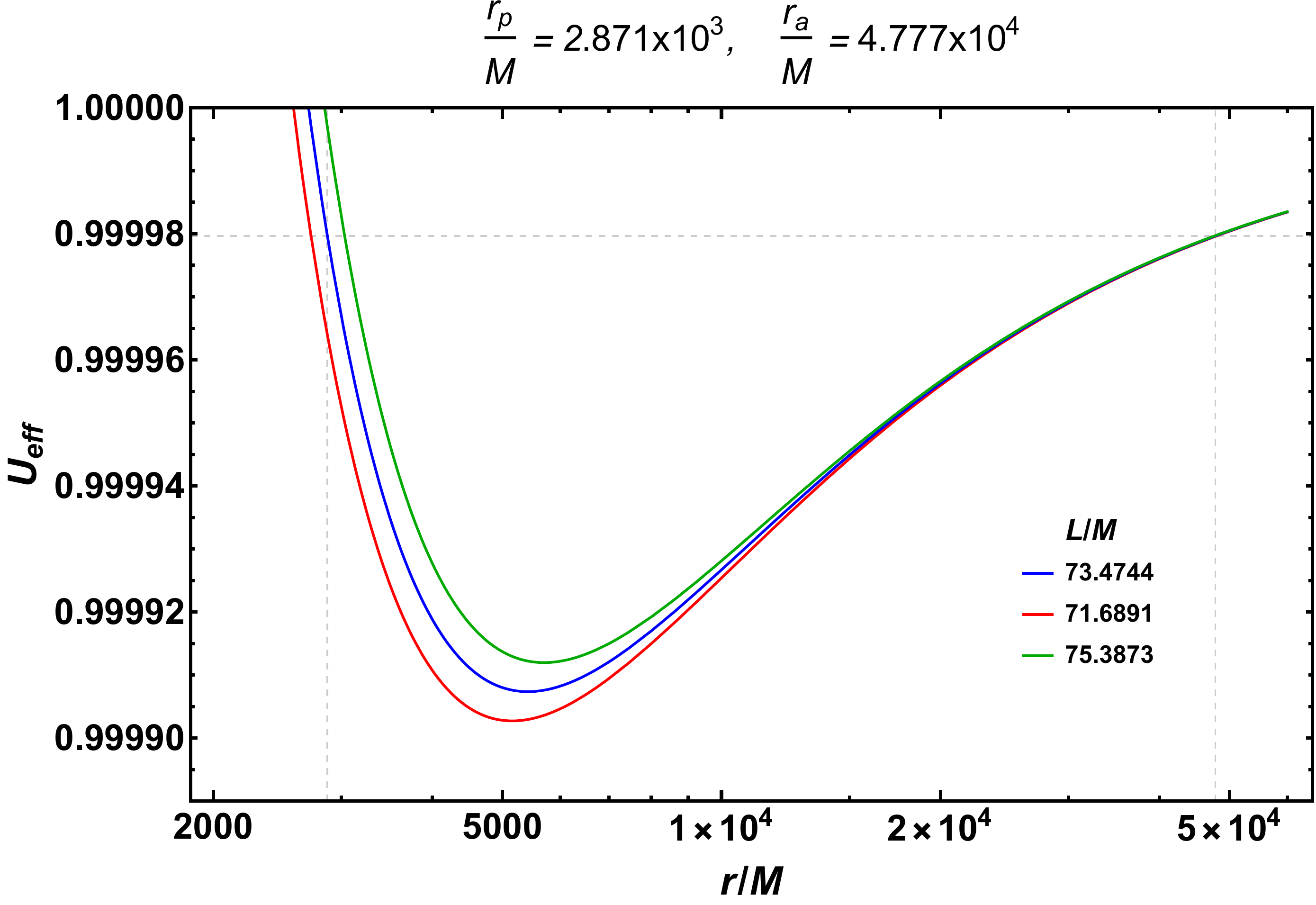}
    \caption{{Effective potential $U_{\rm eff}$ given by \Cref{eqn:Ueff}, for selected values of the conserved angular momentum $L$. Left: massive BH case, i.e. Schwarzschild solution,  $g_{00} = 1-2 M/r$. Right: DM case; $g_{00}$ obtained from numerical integration of the general relativistic equilibrium equations of the extended RAR model for a fermion mass $mc^2=\SI{56}{\kilo\eV}$, see \Cref{sec:SM1} for details. It can be seen that imposing a bound orbit within given values of the pericenter and apocenter (vertical dashed lines), in this example respectively, $r_p/M = 2.976\times 10^3$ and $r_a/M = 4.714\times 10^4$, implies a unique solution of $E$ (dashed horizontal value) and $L$ (value associated with the blue curve). In this example, the adopted mass of the massive BH for the Schwarzschild solution is {$M_{\rm BH} \equiv M= \SI{4.075E6}{\Msun}$, and for the mass of the DM RAR core, has been set to $M_c =\SI{3.5E6}{\Msun}$.}}}
    \label{fig:Ueff}
\end{figure*}

{
First, to solve \Cref{eqn:motiont} we must set the value of $E$ and $L$. From the definition of Lorentz factor, \Cref{eqn:gamma}, and the equation of motion for $t(\tau)$, \Cref{eqn:motiont}, one obtains the first integral:
\begin{equation}\label{eqn:rdotfirstintegral}
    -g_{00}(r) g_{11}(r)\dot{r}^2 = E^2 - U_{\rm eff}^2(r), 
\end{equation}
where 
\begin{equation}\label{eqn:Ueff}
    U^2_{\rm eff}(r) \equiv g_{00}(r)\left(1 + \frac{L^2}{r^2}\right),
\end{equation}
is the well-known \emph{effective potential} governing the radial motion. The relevance of this equation is that it allows to perform a turning-point analysis, analogously to the classical Kepler problem. From \Cref{eqn:rdotfirstintegral}, it can be seen that the request of having a bound, closed orbit within two known turning points, i.e. the pericenter ($r_p$) and the apocenter ($r_a$), where $\dot{r} = 0$, implies a unique solution for $E$ and $L$; see \Cref{fig:Ueff}. The value of $U_{\rm eff}$ at the turning points has to be the same, so we obtain $L$ by solving the algebraic equation:
\begin{equation}\label{eqn:L}
U_{\rm eff}(L,r_p)=U_{\rm eff}(L,r_a),
\end{equation}
and with the knowledge of $L$, we obtain the energy by 
\begin{equation}\label{eqn:E}
E = U_{\rm eff}(L,r_p),\quad {\rm or}\quad E = U_{\rm eff}(L,r_a).
\end{equation}
}

{
The metric functions $g_{00}(r)$ and $g_{11}(r)$ in the BH case are set by the mass of the BH, $M_{\rm BH}$. In the {extended-RAR} model, the parameters $\theta_0$, $\beta_0$, $W_0$ and the fermion mass $m$, are well constrained by the rotation curves of the Galaxy (see \Cref{sec:SM1} and \citealp{2018PDU....21...82A,2019PDU....24100278}, for details). Each possible set of parameters gives a mass of the quantum core, $M_c$ (or alternatively of central density; see \citealp{2019PDU....24100278} for details), so, the metric functions are known once we chose a value of $M_c$ {for given halo boundary conditions in agreement with observables (see \Cref{sec:SM1})}.
}

{
Having set the metric functions (i.e. given $M_c$ in the extended-RAR model or $M_{\rm BH}$ in the BH model), having calculated the values of $E$ and $L$ with given pericenter $r_p$ and apocenter $r_a$ distances (or, alternatively the semimajor axis $a$ and the eccentricity $e$), we can integrate the equations of motion (\ref{eqn:motiont})--(\ref{eqn:motionphi}) giving appropriate initial conditions at initial proper time $\tau_0$. We give them at the apocenter, i.e. we set $t_0 \equiv t(\tau_0) = 0$, $r_0\equiv r(t_0)=r_a$, $\phi_0\equiv \phi (t_0) = \pi$ and $\dot{r}(t_0) = 0$. We integrate numerically the equations of motion via an adaptive integrator based on the fourth-order Runge–Kutta (RKF45) method \citep{Fehlberg1970}. We thus obtain $t(\tau)$, $r(t) = r[\tau(t)]$, $\phi(t) = \phi[\tau(t)]$. We recall that $t$ is coordinate time at emission point, so within our adopted approximation of zero time-delay of the photons; see \Cref{eqn:temtobs}.
}

{
Once the variables of the dynamics of the real orbit have been calculated, we proceed to obtain the orbital elements, $i$, $\omega$, $\Omega$, as well as the constant offsets $X_0$ and $Y_0$, from the request that the predicted orbit, i.e. $X(t)$ and $Y(t)$, \Cref{eqn:XYZ}, and the predicted redshift function $z$, fit the observational values, i.e. respectively $X_{\rm obs}$, $Y_{\rm obs}$ and $z_{\rm obs}$.
}

{
In order to quantify the goodness of the fit, we compute the reduced-$\chi^2$ for each of the observables:
\begin{subequations}
\begin{eqnarray}
\bar{\chi}^2_{X} &=& \frac{1}{N_X-p}\sum_{j=1}^{N_X} \frac{\left[X_{{\rm obs},j}-(X+X_0)\right]^2}{\Delta X_{{\rm obs}, j}^2},\label{eqn:chi2X}\\
\bar{\chi}^2_{Y} &=& \frac{1}{N_Y-p}\sum_{j=1}^{N_Y} \frac{\left[Y_{{\rm obs},j}-(Y+Y_0)\right]^2}{\Delta Y_{{\rm obs}, j}^2},\label{eqn:chi2Y}\\
\bar{\chi}^2_{z} &=& \frac{1}{N_z-p}\sum_{j=1}^{N_z} \frac{\left(z_{{\rm obs},j}-z\right)^2}{\Delta z_{{\rm obs}, j}^2},\label{eqn:chi2z}
\end{eqnarray}
\end{subequations}
where the subscript $j$ indicates the $j$-th data element of the observable $\{X_{{\rm obs}, j},Y_{{\rm obs}, j},z_{{\rm obs}, j}\}$,  $\{\Delta X_{{\rm obs}, j}, \Delta Y_{{\rm obs}, j}, \Delta z_{{\rm obs}, j}\}$ is the associated standard deviation of the $j$-th measurement, $\{N_X, N_Y, N_z\}$ are the number of data elements of the observable, and $p$ is the number of model parameters.
}

{
In order to best match with the observational data at the observational times, which are presented in J2000 convention, we have to perform a time-shift to the theoretical data, $\Delta t$. Therefore, we introduce the new time $t^\prime \equiv t - \Delta t$, i.e., we must calculate $r(t^\prime) = r(t-\Delta t)$, $\phi(t^\prime) = \phi(t-\Delta t)$, etc. Thus, the time-shift $\Delta t$ becomes one of the parameters of the fitting process. Due to the above, eqs.~(\ref{eqn:XYobsXY_X0})--(\ref{eqn:XYobsXY_Y0}) and \Cref{eqn:zobs}, are solved in iterative fashion, by varying $\Delta t$, and calculating the orbital parameters that minimize $\bar{\chi}^2_{X}$, $\bar{\chi}^2_{Y}$, and $\bar{\chi}^2_{z}$, for each value of $\Delta t$. In general, we find that the fit of the redshift function is better than the one of the positions. This occurs both for S2 and G2 since observational data of the position is at times somehow scattered. In any case, besides the individual $\chi^2$ values, we evaluate an \emph{overall} performance of every set of parameters by computing the mean of the $\chi^2$:
\begin{equation}\label{eqn:chimean}
    \langle\chi^2\rangle \equiv \frac{1}{3}\left( \bar{\chi}^2_{X} + \bar{\chi}^2_{Y} + \bar{\chi}^2_{z} \right).
\end{equation}
The values of the model parameters reported in Table~\ref{tab:parameter} correspond to the ones that generate the smallest mean $\langle\chi^2\rangle$ for the range of parameters explored. We also report the individual $\bar{\chi}^2_{X}$, $\bar{\chi}^2_{Y}$, and $\bar{\chi}^2_{z}$. It is important to notice that for different values of the parameters we could obtain a better fit of a specific single observable, e.g. $z_{\rm obs}$. For instance, we found for S2 some set of parameters that yield for $\bar{\chi}^2_{z}$ a value as small as $1.03$, with respect to the value $\bar{\chi}^2_{z} \approx 1.28$ of the set of parameters leading to the smallest $\langle\chi^2\rangle$ (see \cref{tab:parameter}).
}

{
Summarizing, our fitting procedure, for a given core mass $M_c$ of the RAR model, or a BH mass $M_{\rm BH}$ in the massive BH model, performs the following steps:
\begin{enumerate}
    \item 
    Set a value for the eccentricity $e$.
    \item
    Set a value for the semimajor axis $a$.
    \item
    Calculate the pericenter $r_p$ and apocenter $r_a$ for the chosen $e$ and $a$.
    \item 
    Using \Cref{eqn:L} and \Cref{eqn:E}, calculate $L$ and $E$, so to
    integrate the equations of motion (\ref{eqn:motiont})--(\ref{eqn:motionphi}) with initial conditions at apocenter: $t_0 = 0$, $r_0 = r_a$, $\phi_0 = \pi$ and $\dot{r}(t_0) = 0$;
    \item 
    Set a value for the constant time-shift $\Delta t$;
    \item 
    Calculate all quantities of the real orbit at the shifted time $t^\prime = t - \Delta t$, i.e. $r(t^\prime)$, $\phi(t^\prime)$, $\dot{r}(t^\prime)$ and $\dot{\phi}(t^\prime)$.
    \item
    At this stage, the redshift function depends only on the orbital elements $\omega$ and $i$, see \Cref{eqn:zobs}, so we obtain them by minimizing $\bar{\chi}^2_z$, eq.~(\ref{eqn:chi2z}).
    \item
    {We} iterate the above steps $5$--$7$ in an appropriate range of $\Delta t$, calculate the sets $\{\Delta t, \omega,i\}$ leading to each minimum $\bar{\chi}^2_z$, and identify the set leading to the \emph{infimum} $\bar{\chi}^2_z$, i.e. the smallest $\bar{\chi}^2_z$.
    \item
    Set a value of $\Omega$.
    \item
     At this stage, the $X$ position depend only on the offset $X_0$, see eqn.~(\ref{eqn:XYobsXY_X0}), so we obtain it by minimizing $\bar{\chi}^2_X$, eq.~(\ref{eqn:chi2X}).
     \item 
     Likewise, the $Y$ position depend only on the offset $Y_0$, see eqn.~(\ref{eqn:XYobsXY_Y0}), so we obtain it by minimizing $\bar{\chi}^2_Y$, eq.~(\ref{eqn:chi2Y}).
     \item
      {We} iterate the above steps $9$--$11$ in an appropriate range of $\Omega$, calculate the sets $\{\Omega, X_0, Y_0\}$ leading to each minimum of $\bar{\chi}^2_X$ and $\bar{\chi}^2_Y$, and identify the set  leading to the \emph{infimum} of $\bar{\chi}^2_X$ and of $\bar{\chi}^2_Y$, i.e. the smallest $\bar{\chi}^2_X$ and $\bar{\chi}^2_Y$.
     \item 
     Having the smallest values of $\bar{\chi}^2_X$, $\bar{\chi}^2_Y$ and $\bar{\chi}^2_z$, calculate the mean $\langle \chi^2 \rangle$ given by \Cref{eqn:chimean}.
     \item 
     The steps $1$--$13$ are iterated for different values of $e$ and $a$ in some appropriate range.
     \item 
    Identify the best-fit parameters as the ones leading to the smallest $\langle \chi^2 \rangle$.
    \item 
    The steps $1$--$15$ can be repeated for different values of the mass of the DM core $M_c$ in the extended-RAR model, or the BH mass $M_{\rm BH}$ in the central massive BH model.
\end{enumerate}
}

\end{appendix}

\end{document}